\documentclass[letterpaper,11pt]{article}

\pdfoutput=1 

\usepackage{jheppub} 
\usepackage[T1]{fontenc}
\usepackage{lmodern}
\usepackage{amsmath}
\usepackage{amssymb}
\usepackage{graphicx}
\usepackage{xspace}
\usepackage{slashed}
\usepackage{multirow}
\usepackage[normalem]{ulem}

\usepackage{epstopdf}

\newcommand{\nn}{\nonumber} 
\newcommand{\bn}{{\bar n}}

\newcommand{\be}{\begin{equation}}
\newcommand{\ee}{\end{equation}}
\newcommand{\bea}{\begin{eqnarray}}
\newcommand{\eea}{\end{eqnarray}}

\newcommand{\vect}[1]{\mathbf{#1}}
\newcommand{\abs}[1]{\left\lvert #1\right\rvert}

\newcommand{\ket}[1]{\left\lvert #1\right\rangle}

\newcommand{\minus}{\!-\!}

\newcommand{\Lqcd}{\Lambda_{\text{QCD}}}

\newcommand{\as}{\alpha_s}
\newcommand{\MSbar}{\overline{\text{MS}}}

\newcommand{\cO}{\mathcal{O}}
\newcommand{\cM}{\mathcal{M}}

\newcommand{\eq}[1]{Eq.~\eqref{eq:#1}}
\newcommand{\eqs}[2]{Eqs.~\eqref{eq:#1} and \eqref{eq:#2}}
\newcommand{\eqss}[3]{Eqs.~\eqref{eq:#1}, \eqref{eq:#2}, and \eqref{eq:#3}}

\renewcommand{\sec}[1]{Sec.~\ref{sec:#1}}
\newcommand{\ssec}[1]{Sec.~\ref{ssec:#1}}

\newcommand{\fig}[1]{Fig.~\ref{fig:#1}}
\newcommand{\figs}[2]{Figs.~\ref{fig:#1} and \ref{fig:#2}}

\allowdisplaybreaks[1]

\DeclareMathOperator{\Tr}{Tr}

\title{ \Large Predictions for energy correlators probing substructure of groomed heavy quark jets}
\author[a]{Christopher Lee,}
\author[b]{Prashant Shrivastava,}
\author[a]{Varun Vaidya}

\affiliation[a]{Theoretical Division, Group T-2, MS B283, Los Alamos National Laboratory, P.O. Box 1663, \\Los Alamos, NM  87545, USA}
\affiliation[b]{Department of Physics, Carnegie Mellon University, 5000 Forbes Avenue, Pittsburgh, PA 15213, USA} 
\emailAdd{clee@lanl.gov}
\emailAdd{prashans@alumni.cmu.edu}
\emailAdd{vvaidya@lanl.gov}

\abstract{We develop an effective field theory (EFT) framework to perform an analytic calculation for energy correlator observables computed on groomed heavy-quark jets. A soft-drop grooming algorithm is applied to a jet initiated by a massive quark to minimize soft contamination effects such as pile-up and multi-parton interactions. We specifically consider the two-particle energy correlator as an initial application of this EFT framework to compute heavy quark jet substructure. We find that there are different regimes for the event shapes, depending on the size of the measured correlator observable, that require the use of different EFT formulations, in which the quark mass and grooming parameters may be relevant or not.  We use the EFT to resum large logarithms in the energy correlator observable in terms of the momentum of a reconstructed heavy hadron to NLL$'$ accuracy and subsequently match it to a full QCD $\mathcal{O}(\alpha_s)$ cross section, which we also compute. We compare our predictions to simulations in  \textsc{Pythia} for $e^+e^-$ collisions. We find a good agreement with partonic simulations, as well as hadronic ones with an appropriate shape function used to describe nonperturbative effects and the heavy quark hadron decay turned off. We also predict the scaling behavior for the leading nonperturbative power correction due to hadronization. Consequently, we can give a prediction for the energy correlator distribution at the level of the reconstructed heavy hadron. This work provides a general framework for the analysis of heavy quark jet substructure observables.}

\begin{document}

{\flushright LA-UR-18-24853 \\}

\maketitle

\section{Introduction}

Jet substructure observables are playing a significant role in numerous experiments at various energies (e.g.~LHC, RHIC). The focus is on precision standard model measurements, e.g.~\cite{Chatrchyan:2012sn,CMS:2013cda,Aad:2015cua}, as well as searches for new physics, e.g.~\cite{CMS:2011bqa,Fleischmann:2013woa,Pilot:2013bla,TheATLAScollaboration:2013qia}. At the same time, jet substructure measurements are increasingly being used as a probe of the Quark-Gluon Plasma (QGP) medium \cite{Andrews:2018jcm}. QCD jets produced from early stage collisions of beam quarks and gluons from two nuclei play a central role in studying the transport properties of QGP. During their propagation through the hot and dense medium, the interaction between the hard jets and the colored medium will lead to parton energy loss (jet quenching) \cite{Bjorken:1982tu,Gyulassy:1990ye,Wang:1991xy}. There have been several experimental signatures of jet energy loss observed at RHIC and the LHC such as modification of reconstructed jets \cite{Aad:2014bxa,Chatrchyan:2012gt,Chatrchyan:2011sx,Aad:2010bu} and jet substructure \cite{Chatrchyan:2012gw,Chatrchyan:2013kwa,Aad:2014wha} as compared to the expectations from proton-proton ($pp$) collisions. Continued progress relies on achieving a deeper understanding of the dynamics of jets, allowing for subtle features in a jet to be exploited. This understanding has progressed rapidly in recent years, both due to advances in explicit calculations, e.g.~\cite{Feige:2012vc,Field:2012rw,Dasgupta:2013ihk,Dasgupta:2013via,Larkoski:2014pca,Dasgupta:2015yua,Li:2017wwc},  as well as due to the development of techniques for understanding dominant properties of substructure observables using analytic \cite{Walsh:2011fz,Larkoski:2014gra,Larkoski:2014zma} and machine learning \cite{Cogan:2014oua,deOliveira:2015xxd,Almeida:2015jua,Baldi:2016fql,Guest:2016iqz,Conway:2016caq,Barnard:2016qma} approaches.  Developments in jet substructure (see \cite{Larkoski:2017jix} for a recent review) have shown that the modified mass drop tagging algorithm (mMDT) or soft-drop grooming procedure robustly removes contamination from both underlying event and non-global color-correlations, see Refs. \cite{Larkoski:2014wba,Dasgupta:2013via,Dasgupta:2013ihk}, and have been applied to study a wide variety of QCD phenomenology within jets \cite{Larkoski:2017iuy,Hoang:2017kmk,Marzani:2017mva,Dasgupta:2016ktv,Frye:2016okc,Frye:2016aiz,Larkoski:2015lea,Dasgupta:2015lxh,Chien:2014nsa}.

Heavy $b$ quark jets play a particularly prominent role in QGP tomography. It was pointed out in \cite{Dokshitzer:2001zm}  that heavy quarks have fundamentally different radiation patterns from light quarks and thus heavy quark jets are expected to be more greatly affected by plasma interactions. Experiments at RHIC \cite{Frawley:2008kk} and LHC take advantage of these properties to make extensive studies of $b$ jets, and the planned sPHENIX experiment in particular promises improved tracking and flavor tagging capabilities for higher precision measurements on $b$ jets \cite{Adare:2015kwa}. Thus improved theory calculations of $b$ jet properties are now very timely. 

In this paper we take steps to build and apply a theoretical framework to compute the substructure of jets initiated by heavy quarks, starting in vacuum. Much of the necessary framework has already been developed in the context of computations of jet mass for ungroomed \cite{Fleming:2007qr,Fleming:2007xt} and groomed \cite{Hoang:2017kmk} top quark jets as well as transverse momentum spectra for $B$ hadrons \cite{Makris:2018npl}. We can also build upon the EFT framework for groomed light jets in \cite{Frye:2016aiz}. Meanwhile \cite{Pietrulewicz:2017gxc} built an extensive EFT framework for massive quark effects on measurements in Drell-Yan.  We draw upon many of these analyses, organizing their pieces to be applied to a large class of Infrared and Collinear (IRC) safe event shapes known as energy correlation functions \cite{Banfi:2004yd,Jankowiak:2011qa,Larkoski:2013eya} computed for groomed massive quark jets. We find the appropriate EFT to describe these observables for groomed massive quark jets is an elegant combination of those for ungroomed massive jets in \cite{Fleming:2007qr,Fleming:2007xt} and groomed light-parton jets in \cite{Frye:2016aiz}, resulting in an EFT very similar to that used in  \cite{Hoang:2017kmk} for groomed top jets. We use this EFT framework to smoothly describe the whole spectrum in the 2-point energy correlator called $e_2^{(\alpha)}$:
\begin{itemize}
\item For very large $e_2^{(\alpha)}$, the spectrum is predicted by fixed-order perturbation theory in full QCD, to which the predictions of resummed singular terms in $e_2^{(\alpha)}$ for smaller values will have to matched.

\item For intermediate $e_2^{(\alpha)}$, the spectrum is sensitive to soft drop grooming effects, but the radiation is sufficiently energetic or at wide enough angles not to be sensitive to quark mass. The $e_2^{(\alpha)}$ dependence in the spectrum factorizes into a \emph{collinear-soft} mode \cite{Bauer:2011uc} and a \emph{collinear (jet)} mode equivalent to that for massless parton jets.\footnote{This region may or may not exist separately, depending on the size of the angular exponent $\alpha$ used in $e_2^{(\alpha)}$.}

\item For smaller $e_2^{(\alpha)}$, the spectrum is sensitive to radiation at small enough angles to be affected by the nonzero quark mass. The $e_2^{(\alpha)}$ dependence in the spectrum factorizes into the aforementioned collinear-soft mode, a matching coefficient to HQET for a heavy quark of mass $m_b$, and an \emph{ultracollinear} mode \cite{Fleming:2007qr,Fleming:2007xt} describing soft radiation from the $b$ quark in its rest frame that is highly boosted into the lab frame, where it looks collinear.

\item For very small $e_2^{(\alpha)}$, we encounter a nonzero minimum cutoff $e_2^{(\alpha)}\geq e_{2,\text{min}}^{(\alpha)}$ imposed by grooming and the finite quark mass. Near this scale, we find the collinear-soft and ultracollinear modes actually merge back into each other, recombining into a single groomed massive quark jet function describing the full $e_2^{(\alpha)}$ dependence here, and requires a fixed-order matching from the region above it, something not usually required at \emph{small} values of a jet shape observable. This is a novel feature of applying the EFTs developed in \cite{Frye:2016aiz,Hoang:2017kmk} that we find for this particular class of observables with massive quark jets.
\end{itemize}
In the main part of the paper we will explain each of these regions in detail. See in particular \fig{ztheta} and \fig{ztheta2} for an intuitive illustration of how these modes arise in each region. We see that it is primarily in the last two regions that the heavy quark mass affects the distribution, causing it to differ from those for light parton jets.

The EFT developed here will thus reveal general properties of the radiation pattern in a massive quark jet as well as the impact of the quark mass in determining the jet substructure. In this paper we apply the EFT to compute energy correlation functions for heavy quark jets produced in $e^+e^-$ collisions, as a first step towards computations for $pp$ collisions, and ultimately, heavy-ion collisions. The act of grooming, however, should help isolate properties intrinsic to the heavy quark jet itself, reducing differences between $e^+e^-$ and $pp$ cases, at least in the shape of the energy correlator distributions. Grooming also allows $b$ jets to be probed closer to their mass threshold $m_J^2 \gtrsim m_b^2$ in a way that can still be analyzed perturbatively with controlled nonperturbative corrections by removing the wide-angle soft radiation that in the ungroomed case masks this region, and makes boosted HQET practically applicable to $b$ jets for the first time (cf. \cite{Dehnadi:2016snl,Hoang:2019fze}). An analytic understanding of the jet substructure properties for heavy quark-initiated jets in  collisions in vacuum will serve as a comparison baseline for understanding the modification of the properties by the medium. Here, we focus just on two-point energy correlation functions, but the EFT that we develop here is quite general and can also be applied to other jet substructure observables computed on heavy quark jets.

At the same time, the hope is that by understanding the decay kinematics of heavy quark hadrons, jet substructure observables can ultimately serve as a complementary technique to heavy quark jet tagging. Generalized energy correlators have been shown to be excellent discriminants of light quark~vs.~gluon jets \cite{Moult:2016cvt}, and we have performed preliminary studies showing promising performance for heavy~vs.~light jet discrimination. This application will require the computation of energy correlation functions in terms of the light decay products of heavy hadrons, which goes beyond the scope of this paper. Here we focus instead on two-point energy correlator functions in terms of the momentum of a reconstructed heavy hadron, assuming it has been identified by other means, such as displaced vertex tracking. We leave the extension to include the effects of $B$ hadron decay to a future publication.

In \sec{quarkSubstructure}, we introduce the specific class of observables that we wish to use to study heavy quark jet substructure.
In \sec{factorization}, we give details of the factorization theorem (formulated within SCET and HQET) for resumming large logarithms in our jet observable. \sec{oneloopEFT} gives details about the one loop EFT calculation of the factorization ingredients and anomalous dimensions used for resummation. \sec{fixed} discusses the numerical implementation of the full theory fixed-order calculation at $\cO(\alpha_s)$ used to match in the far-tail region of the energy correlator distributions, along with an analytical calculation of the singular limit. \sec{resum} provides the analytical expression for the resummed result to NLL accuracy. We then compare our resummed result matched to the $\cO(\alpha_s)$ full theory fixed-order cross section with \textsc{Pythia} at both parton level in \sec{resum} and hadron level in \sec{decayq}. In \sec{decayq} we also discuss the anticipated impact of $B$ hadron decay on the shape of the distribution. Here we also give a prediction for the scaling of the leading nonperturbative corrections using the energy flow operator technique. We conclude in \sec{conclusion}. 


\section{Probing substructure of heavy quark jets}
\label{sec:quarkSubstructure}
In this section we describe the jet substructure observables that are our focus in this paper. We also provide a review of the soft drop grooming algorithm that we use throughout this paper. The general class of substructure observables that we consider  fall in the class known as IRC safe observables. This means that these observables are insensitive to arbitrarily soft or collinear radiation or splitting. Of course, in case of heavy quark initiated jets, the mass of the heavy quark provides a cutoff on the collinearity of the radiation. This, combined with grooming (which restricts the softness of the radiation), restricts the minimum value achievable for any jet substructure observable. This is one of the general features that distinguishes massive quark jets from massless ones. 
\subsection{Energy Correlation Functions}
\label{Energy_corr}
We focus on the measurement of the two-point energy correlation function \cite{Larkoski:2013eya,Jankowiak:2011qa} on a jet initiated by a heavy quark. The jet is groomed using a soft drop grooming algorithm.
For $e^+ e^-$ collisions this observable is defined as 
\bea
e_2^{(\alpha)}|_{e^+e^-} =  \sum_{i<j \in J} z_i z_j (\theta_{ij})^{\alpha} 
\label{epem}
\eea
where we sum over pairs of particles $(i,j)$ inside the groomed jet. We define the angle ($\theta_{ij}$)  between the two particles with momentum $p_i$ and $p_j$ as 
\bea
\label{eq:eeangle}
\theta^2_{ij} =  \frac{2 p_i \cdot p_j}{E_i E_j} 
\eea
while the energy fractions are $z_i = E_i/E_J$, $E_J$ being the energy of the jet. The energy correlation functions are insensitive to recoil effects \cite{Banfi:2004yd,Larkoski:2014uqa}. At the same time, they do not include explicit axes in their definition.

In the small angle and massless limits, \eq{eeangle} coincides with the geometric angle between the three-vectors $\vect{p}_i$ and $\vect{p}_j$, which we will denote $\vartheta_{ij}$ when we need to refer to it:
\begin{align}
\label{eq:geometric}
\theta_{ij}^2 = 2\frac{E_i E_j - \abs{\vect{p}_i}\abs{\vect{p}_j}\cos\vartheta_{ij}}{E_i E_j} &= 2\Bigl(1 - \sqrt{1-m^2/E^2_i}\cos\vartheta_{ij} \Bigr) \\
&= \vartheta_{ij}^2 + \frac{m^2}{E_i^2}  + \cO(\vartheta^4, m^2 \vartheta^2 / E^2) \,, \nn
\end{align}
where we took one of the particles ($i$) to have a nonzero mass. In the small mass and small angle regimes, we can drop the corrections, as appropriate. \eq{geometric}, does, however, imply a nonzero minimum for the $\theta$ between a heavy quark and soft gluon radiation it emits, for which we can let $\vartheta$ go all the way to zero:
\be
\label{eq:mintheta}
\theta_{gb}\geq \theta_{\text{min}} = \frac{m}{E_J}\,,
\ee
an effect we will find important to keep.

For jets produced in $pp$ collisions, the definition of the energy correlation function is modified to be boost invariant along the beam direction.
\bea
e_2^{(\alpha)}|_{pp} = \frac{1}{p^2_{T_J}} \sum_{i<j\in J}p_{T_i}p_{T_j}R^{\alpha}_{ij},
\label{pp}
\eea
where $p_{T_J}$ is the transverse momentum of the jet, $R_{ij}$ is defined as, 
\bea
 R^2_{ij}= (\phi_i-\phi_j)^2+(y_i-y_j)^2,
\eea
and $p_{T_i},\phi_i$ and $y_i$ are the transverse momentum, azimuthal angle and rapidity of particle $i$, respectively. In this paper, we focus on the $e^+e^-$ calculation. However, it was shown in \cite{Frye:2016aiz}, that for jets at central rapidities and in the limit that all the particles in the jet are collinear, the two definitions of the energy correlator function in Eq.~(\ref{epem}) and Eq.~(\ref{pp}) are equivalent. 

\subsection{Soft Drop Grooming algorithm}
\label{ssec:soft-drop}
The modified mass-drop procedure (mMDT) \cite{Dasgupta:2013via,Dasgupta:2013ihk} or its generalization known as soft drop \cite{Larkoski:2014wba} removes contaminating soft radiation from a jet, which is originally identified by any suitable algorithm such as (anti-)$k_t$ \cite{Catani:1991hj,Catani:1993hr,Ellis:1993tq,Dokshitzer:1997in,Cacciari:2008gp}, by reconstructing an angular ordered tree of the jet constituents by using the Cambridge/Aachen (C/A) clustering algorithm \cite{Ellis:1993tq,Catani:1993hr,Dokshitzer:1997in,Wobisch:1998wt,Wobisch:2000dk}, and removing the branches at the widest angles which fail an energy requirement. As soon as a branch is found that passes the test, it is declared the groomed jet, and all the constituents of the branch are the groomed constituents. One simply finds the branch whose daughters are sufficiently energetic. Formally the daughters could have any opening angle, though their most likely configuration is collinear.

The strict definition of the algorithm is as follows. Given an ungroomed jet, first we build the clustering history by starting with a list of particles in the jet. At each stage we merge the two particles within the list that are closest in angle\footnote{This merging is usually taken to be summing the momenta of the particles, though one could use winner-take-all schemes \cite{Salam:WTAUnpublished,Bertolini:2013iqa,Larkoski:2014uqa}.}. This gives a pseudo-particle, and we remove the two daughters from the current list of particles, replacing them with the merged pseudo-particle. This is repeated until all particles are merged into a single parent. Then we open the tree back up. At each stage of the declustering, we have two branches available, label them $i$ and $j$. We require:
\begin{align}
\label{SD:condition}
\frac{\text{min}\{E_i,E_j\}}{E_i+E_j}>z_\text{cut}\left(\frac{\theta_{ij}}{R}\right)^{\beta},
\end{align}
where $z_\text{cut}$ is the modified mass drop parameter, $\beta$ is the parameter which controls the angularities, $\theta_{ij}$ is the angle between $i^\text{th}$ and $j^\text{th}$ particle defined in \eq{eeangle}, $R$ is the jet radius and $E_i$ is the energy of the branch $i$. (In this paper we will actually just stick to $\beta=0$, for which soft drop coincides mMDT.) If the two branches fail this requirement, the softer branch is removed from the jet, and we decluster the harder branch, once again testing Eq.~(\ref{SD:condition}) within the hard branch. The pruning continues until we have a branch that when declustered passes the condition Eq.~(\ref{SD:condition}). All particles contained within this branch whose daughters are sufficiently energetic constitute the groomed jet. Intuitively we have identified the first genuine collinear splitting.

For a hadron-hadron collision, one uses the transverse momentum $(p_T)$ with respect to the beam for the condition of Eq.~(\ref{SD:condition}),
\begin{align}\label{SD:condition_pp}
\frac{\text{min}\{p_{Ti},p_{Tj}\}}{p_{Ti}+p_{Tj}}>z_\text{cut}\left(\frac{\theta_{ij}}{R}\right)^{\beta}.
\end{align}

We formally adopt the power counting $z_\text{cut} \ll 1$, though typically one chooses $z_\text{cut} \sim 0.1$. See \cite{Marzani:2017mva} for a study on the magnitude of the power corrections with respect to $z_\text{cut}$ for jet mass distributions.


\section{Factorization in SCET$_+$ and bHQET}
\label{sec:factorization}

We work within the formalism of SCET (Soft Collinear Effective Theory) \cite{Bauer:2000yr,Bauer:2001ct,Bauer:2001yt,Bauer:2002nz,Beneke:2002ph}, which provides an EFT framework for studying IR modes in jet physics. Due to the presence of a heavy quark, we will also need HQET (Heavy Quark Effective Theory) \cite{Manohar:2000dt,Isgur:1989vq,Isgur:1989ed}, which is an EFT with an expansion parameter $1/m_q$, the inverse of heavy quark mass, or more specifically, boosted HQET (bHQET) \cite{Fleming:2007qr,Fleming:2007xt} due to large energy of the $b$ quark jet. Due to additional scales induced by the jet grooming we will also need to use the extension of SCET known as SCET$_+$ \cite{Bauer:2011uc,Pietrulewicz:2016nwo}.

We will predict the cross section in $e_2^{(\alpha)}$ defined in Eq.~(\ref{epem}) measured on jets initiated by a heavy quark to which soft drop grooming has been applied. Schematically, the prediction for this spectrum takes the form
\be
\label{eq:singns}
\frac{1}{\sigma_0}\frac{d\sigma}{de_2^{(\alpha)}} = \sigma_\text{sing}(e_2^{(\alpha)}) + \sigma_\text{ns}(e_2^{(\alpha)})\,,
\ee
where $\sigma_\text{sing}$ contains logs of $e_2^{(\alpha)}$, which become large for $e_2^\alpha\ll 1$ and which we will resum to give an accurate prediction in this region. It is to $\sigma_\text{sing}$ that the factorization in this section applies. Meanwhile, $\sigma_\text{ns}$ is the nonsingular (that is, integrable as $e_2^{(\alpha)}\to 0$) remainder function that matches the resummed singular prediction to the prediction of fixed-order perturbation theory in full QCD, which is accurate for larger $e_2^{(\alpha)}$. We will compute $\sigma_\text{ns}$ in \sec{fixed} and discuss how we smoothly interpolate between the two terms in \eq{singns} in \sec{resum}.

The precise set of EFT modes that are needed to factorize $\sigma_\text{sing}$ depends, as we will explain, on the relative hierarchy of scales amongst $e_2^{(\alpha)},z_\text{cut}$ and $m_q$. In the rest of this section we will consider these various possible hierarchies, identify the appropriate modes, and provide a factorized form of $\sigma_\text{sing}$.


\subsection{Power counting and modes}
\label{modes}
 An efficient approach for studying jet substructure in a systematic fashion is power counting \cite{Walsh:2011fz,Larkoski:2014gra}, which allows us to determine the parametric scaling of observables. This is determined by the soft and collinear limit of QCD and is a very powerful technique to determine the structure of factorization.  We want to calculate the two-point energy correlation function Eq.~(\ref{epem}) in $e^+e^-$ collisions on a massive quark jet which we have identified a jet using an appropriate jet algorithm such as anti-$k_t$ or Cambridge/Aachen, and which  is then groomed with a soft drop grooming algorithm described in \ssec{soft-drop}. This makes it relatively insensitive to recoil effects and, in the analogue in hadron collisions, to MPI (multi-parton interactions). At the same time, the effect of non-global logarithms is negligible \cite{Frye:2016aiz}. The grooming parameters are $z_\text{cut}$ and $\beta$, and we will exclusively work with $\beta=0$ for which soft drop coincides with mMDT. A typical value of $z_\text{cut}$ used is $\sim 0.1$.

We consider a jet with energy $E_{J}$ and radius $R\sim\mathcal{O}(1)$ in power counting, so we do not consider resumming logs of $R$, which could be done using additional modes as in \cite{Chien:2015cka,Becher:2015hka,Becher:2016mmh,Kolodrubetz:2016dzb}. For the computation of $\sigma_\text{sing}$, we are working in a regime in where $e_2^{(\alpha)} \ll z_\text{cut}$. (At $e_2^{(\alpha)}\sim z_\text{cut}$, we transition to the fixed-order $\sigma_\text{ns}$ in \eq{singns} and computed in \sec{fixed}.) We also have a hierarchy $m_q \ll E_J$, so that the heavy quark is highly boosted. The jet axis is defined by the light-like $n$ direction, defined as $n= (1,\vect{\hat z})$ in Minkowski coordinates, where $\vect{\hat z}$ is a unit 3-vector in the direction of the jet. 

\begin{figure}
\centerline{\scalebox{.72}{\includegraphics{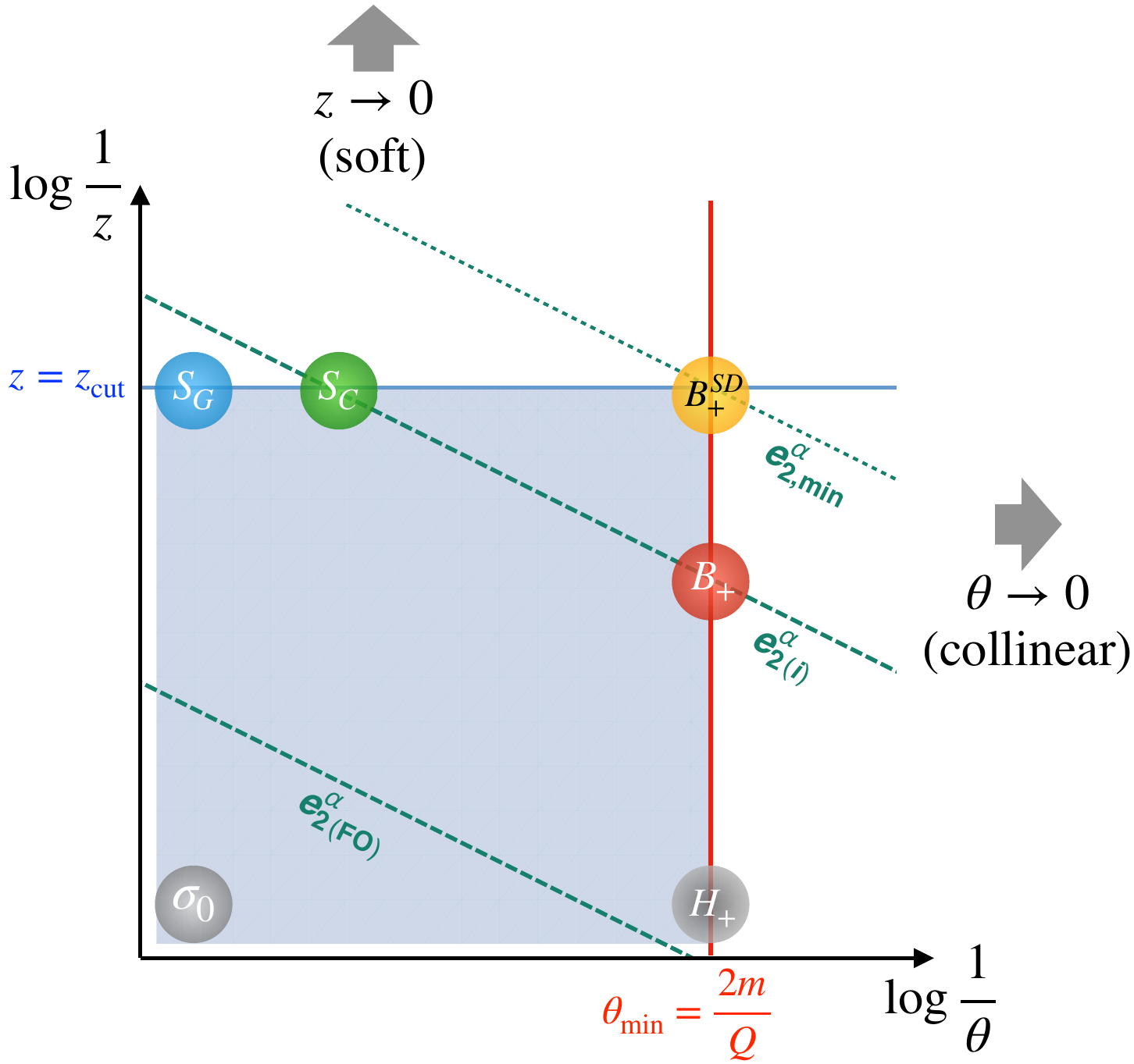}}}
\vskip-0.3cm
\caption[1]{Phase space in $z,\theta$ and associated modes in the singular limit. The allowed phase space is determined by imposing the soft drop cut on the energy $z_\text{cut}<z$, and the limit on the angle $\theta$ imposed by the jet radius, $\theta<R$ and the finite quark mass $\theta>\theta_\text{min} = m/E_J$. Lines of constant $e_2^{(\alpha)}$ given by \eq{e2line} are shown at several values of $e_2^{(\alpha)}$. The light dotted line is the minimum value $e_{2,\text{min}}^{(\alpha)}$ in \eq{e2min}. The first dashed line at $e_{2(i)}^\alpha$ represents a small value where the cross section is sensitive both to the quark mass and to grooming. The second dashed line at $e_{2\text{(FO)}}^\alpha$ represents a larger value where the cross section is not sensitive to quark mass or grooming and has to computed in fixed-order perturbation theory in full QCD (the exact shape of the phase space boundaries at larger $z,\theta$ will also have to be taken into account, see \sec{fixed}). Note that this plot is made for the case $\alpha = \frac{1}{2}<1$. (See \fig{ztheta2} for the case $\alpha>1$.) At $e_{2,\text{min}}^{(\alpha)}$, the collinear-soft and ultracollinear modes merge back into a single-scale, soft drop-sensitive ultracollinear mode ($B_+^{SD}$). The relevant regions for the global soft function $S_G$, the hard function $\sigma_0$, and the matching coefficient $H_+$ to bHQET are also illustrated. }
\label{fig:ztheta} 
\end{figure}

Let us see what the measurement of $e_2^{(\alpha)}$ on the constituents of a groomed jet implies for the momentum scaling of relevant degrees of freedom in the final state. We can divide the modes within the groomed jet into two categories, those which are at a wide angle $\theta_s \sim 1$ relative to the jet axis and those which are collinear to it, $\theta_c\ll 1$. In the following discussion it will be useful to make reference to \fig{ztheta} to see where the relevant modes appear. This figure is analogous to Fig.~2 in \cite{Frye:2016aiz} for massless parton jets, to which the new ingredients due to the heavy quark mass appear along the line at $\theta=\theta_\text{min}$ in \fig{ztheta}. The plot is in $\log(1/\theta)$ and $\log(1/z)$, where $z$ and $\theta$ are the fraction of the jet energy $z=E/E_J$ and angle from the heavy quark $\theta_{ib}$ defined by \eq{eeangle} carried by the radiation from the $b$ quark. In these variables, in the soft and collinear limits, the phase space constraints imposed by soft drop, the jet radius, and the minimum angle due to the nonzero quark mass, are all simple straight lines, and contours of constant $e_2^{(\alpha)}$ are also straight lines,
\be
\label{eq:e2line}
\log \frac{1}{z} = \log\frac{1}{e_2^{(\alpha)}} - \alpha \log\frac{1}{\theta}\,.
\ee
The precise hierarchies of modes that appears depends on the size of $\alpha$; the case $\alpha<1$, which will be the main focus of our paper, is shown in \fig{ztheta}. We will include the discussion of $\alpha>1$ in what follows in this section; the analogous diagram of modes is in \fig{ztheta2}.

\subsubsection*{Wide-angle soft modes}
First we consider the wide angle radiation, $\theta_s\sim 1$.
Assuming that the heavy quark in the jet carries most of the energy (i.e. $z\sim1$), the contribution of this wide-angle radiation with energy fraction $z_s$ to the measurement of $e_2^{(\alpha)}$ would be just $e_2^{(\alpha)} \sim z_s$. For measured values $e_2^{(\alpha)}\ll z_\text{cut}$, which holds in the singular region we are considering, this implies $z_s\ll z_\text{cut}$, so any such radiation that could contribute to the measurement of $e_2^{(\alpha)}$ would be groomed away. The grooming acts effectively as a veto on soft radiation with $z_s< z_\text{cut}$ and contributes to the normalization of the cross section through a soft function $S(E_J z_\text{cut},R)$, but does not affect the shape of the $e_2^{(\alpha)}$ distribution.  This soft function describes the contribution of modes whose momentum $(p_s)$ scales, in light-cone coordinates $p = (\bar n\cdot p, n\cdot p, \vect{p}_\perp)$, as
\bea
p_s \equiv E_J z_\text{cut} (1, 1, \vect{1}). 
\eea
The energy of this mode indicates that it is sensitive to the $z_\text{cut}$ parameter but is groomed away since it fails soft drop. This mode is unable to resolve any smaller angle structure induced by a measurement of $e_2^{(\alpha)}\ll z_\text{cut}$ (e.g. $e_{2(i)}^{(\alpha)}$ in \fig{ztheta}) or by the quark mass $m_b$.
For larger $e_2^{(\alpha)}$ (e.g. $e_{2(\text{FO})}^{(\alpha)}$ in \fig{ztheta}), the wide-angle radiation could contribute to the measurement; however, in this region, the distribution must be calculated in fixed-order perturbation theory, i.e. $\sigma_\text{ns}$ in \eq{singns}, which we will calculate in \sec{fixed}.\footnote{Note that without grooming, wide-angle radiation does contribute to measurement of $e_2^{(\alpha)}$. When this radiation is nonperturbative, $p_s\sim \Lqcd(1,1,\vect{1})$, and the resulting jet mass is $m_J^2\sim Q\Lqcd\gg m_b^2$ for typical $Q$. Then we could never reach the lower jet scales including the bHQET region identified below \cite{Dehnadi:2016snl,Hoang:2019fze}. Grooming is what allows $b$ jets to be probed closer to the heavy quark mass threshold.}

\begin{figure}
\centerline{\scalebox{.7}{\includegraphics{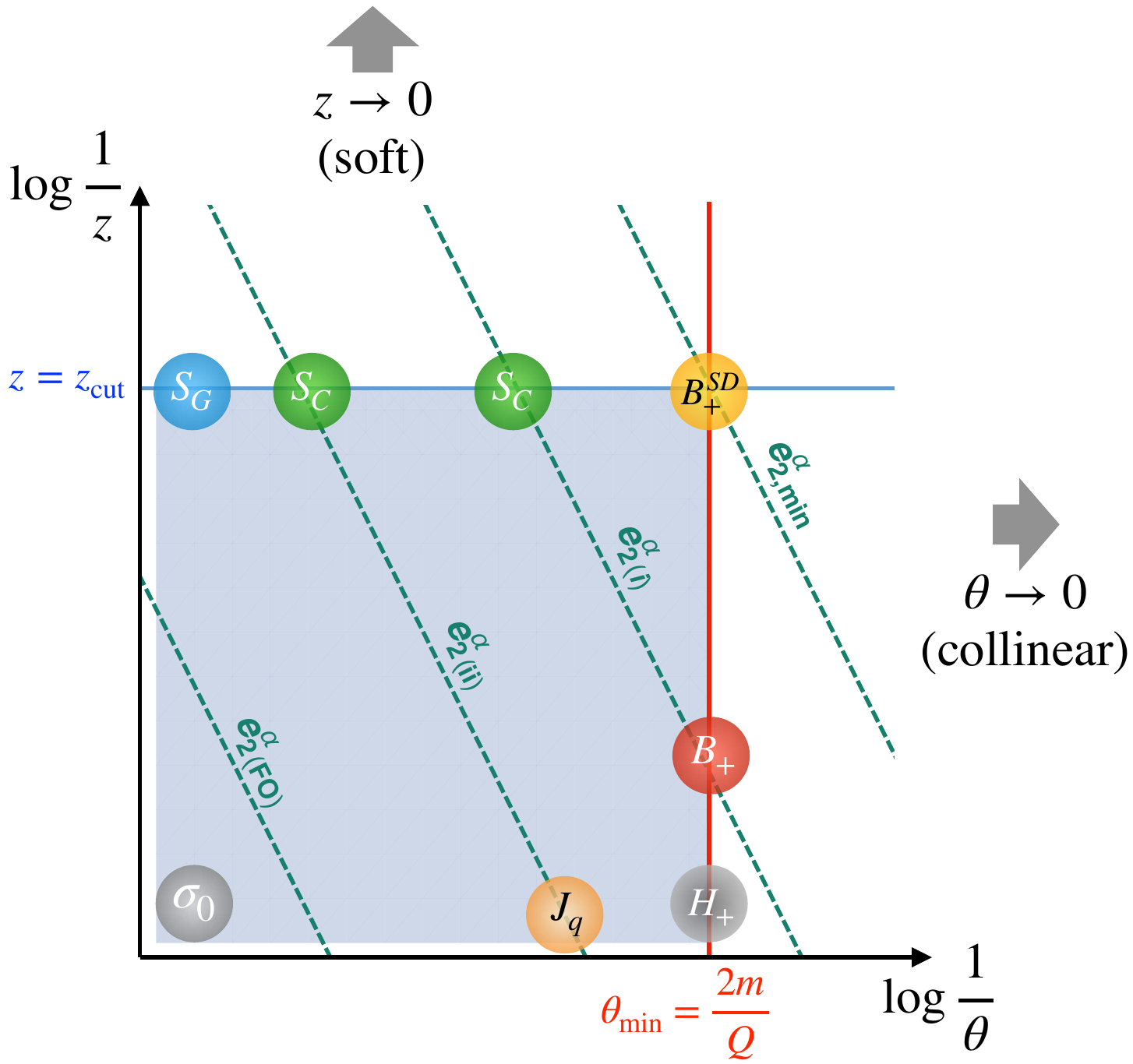}}}
\vskip-0.3cm
\caption[1]{Phase space in $z,\theta$ and associated modes, for the case $\alpha >1$ (in this plot, $\alpha=2$), to be contrasted with \fig{ztheta}. In this case, the $e_2^{(\alpha)}$ spectrum for values represented by $e_{2(ii)}^{(\alpha)}$ is insensitive to the quark mass, and the collinearity of the jet modes is determined solely by $e_2^{(\alpha)}$ itself. In this region, the factorization is the same as for massless parton groomed jets, giving rise to the massless jet function $J_q$. For smaller $e_2^{(\alpha)}$ values represented by $e_{2(i)}^{(\alpha)}$, the factorization is the same as in \fig{ztheta}.}
\label{fig:ztheta2} 
\end{figure}

The factorization of the cross section at this stage then yields the same form as for massless groomed jets in \cite{Frye:2016aiz}, but with a replacement of the massless jet function by a massive jet function:
\bea
\label{eq:factorization1}
\frac{d \sigma}{d e_2^{(\alpha)}} = \sigma_0( Q^2,R,\mu) \times S\left(E_Jz_\text{cut},\mu\right) \times J_{qz}\left( e_2^{(\alpha)}, m_q/E_J, z_\text{cut},\mu\right), 
\eea
where $J_{qz}$ is a massive quark jet function that contains all the radiation in the jet that may contribute to the measurement and $S(E_Jz_\text{cut})$ is the soft function. The jet function $J_{qz}$ is still sensitive to multiple scales, $m_q$, the mass of the heavy quark, the jet energy $E_J$ and  the grooming parameter $z_\text{cut}$. For the range of values $e_{2,\text{min}}^{(\alpha)} \ll e_2^{(\alpha)}$ and $z_\text{cut}$ that we will be interested in, there is still a wide scale separation within this function, as we are about to discuss below, and is already illustrated along the $e_{2(i)}^{(\alpha)}$ contour in \fig{ztheta} or the $e_{2(i,ii)}^{(\alpha)}$ contours in \fig{ztheta2}. It thus requires further factorization. We will give a formal definition at these later steps of factorization.

The factor $\sigma_0$ in \eq{factorization1} contains the Born cross section, the hard matching function $H(Q^2,\mu)$, and an unmeasured jet function $J_\bn(QR,\mu)$ \cite{Chien:2015cka,Ellis:2009wj,Ellis:2010rwa} for the jet in the opposite direction on which $e_2^{(\alpha)}$ is not measured, as well as another soft factor for radiation between the two jets (which disappears in the case of hemisphere jets $R\to 1$). All of these factors describe effects at much higher energy scales that we integrate out of our EFT, represented by the bottom left corner of the phase space in \fig{ztheta}, and just contribute to the normalization but not the shape of the $e_2^{(\alpha)}$ distribution; they will all be divided out later.

%


\subsubsection*{Hierarchy of collinear modes}

We next consider the collinear radiation. Due to the grooming and the quark mass, there are multiple ``collinear'' scales.
The first important observation to make is that there is a minimum angle $\theta_{ij}$ as defined in \eq{eeangle} that can exist between collinear radiation and the massive quark that initiated it, due to nonzero $m_q$. For the case of massless jets there is no such lower bound. Hence, in such a jet, for a mode with a given energy, the angle is set by the measurement of $e_2^{(\alpha)}$ alone. On the other hand, for a jet initiated by a massive quark,  $\theta_{ij}$ between the heavy quark and a light parton has a minimum value as we observed in \eq{mintheta}:
\bea
\label{eq:thetamin}
\theta_\text{min} = \frac{m_q}{E_J} +\mathcal{O}\left(\frac{m_{q}}{E_{J}}\right)^2,
\eea
which comes from the condition $\vartheta_{ij} \geq 0$ for the \emph{geometric} angle, and $E_i\leq E_J$ for the quark energy, in \eq{geometric}.
Since the minimum energy fraction of a light parton in the jet to pass soft-drop needs to be $z_\text{cut}$, this automatically sets a lower limit on the values of $e_2^{(\alpha)}$,
\bea
\label{eq:e2min}
e_{2,\text{min}}^{(\alpha)} = z_\text{cut} \left(\frac{m_q}{E_J}\right)^{\alpha}+ \text{power corrections}, 
\eea
which is evident as the value where the $e_2^{(\alpha)}$ contours in \fig{ztheta} and \fig{ztheta2} exit the allowed phase space set by grooming and by the quark mass. 
Depending on the hierarchy of scales, the angular scaling of a mode is then set by either the measurement of $e_2^{(\alpha)}$ alone or by $\theta_\text{min}$. 
For different regions of $e_2^{(\alpha)}$, then, the relevant collinear degrees of freedom depend on whether they are sensitive to this minimum cutoff or not. Let us systematically go through the possible hierarchies.

\paragraph{Collinear-soft modes} Consider first, the widest angle radiation which we could have in the groomed jet. To pass grooming, it must have an energy fraction $z \gtrsim z_\text{cut}$. Since the jet is measured to have a value of $e_2^{(\alpha)}$, this radiation must have an angle scaling as
\bea
\label{eq:thetacs}
\theta \sim \left(\frac{e_2^{(\alpha)}}{z_\text{cut}}\right)^{1/\alpha}. 
\eea
Given the lower limit \eq{e2min}, on $e_2^{(\alpha)}$, we immediately see that this angle satisfies
\bea
\theta \geq \theta_\text{min} \,.
\eea
Hence the angular scaling will be determined by the measured $e_2^{(\alpha)}$ as in \eq{thetacs}, and the light-cone components of the momentum of this mode must scale as 
\bea
\label{eq:pcs}
p_{cs} \sim z_\text{cut} E_J\left( 1, \left(\frac{e_2^{(\alpha)}}{z_\text{cut}}\right)^{2/\alpha}, \left(\frac{e_2^{(\alpha)}}{z_\text{cut}}\right)^{1/\alpha}\right) \,,
\eea
which, as we have labeled, is the scaling of a \emph{collinear-soft} mode \cite{Bauer:2011uc}. It has a soft enough energy to be sensitive to grooming at $z_\text{cut}$, and wider angle than the ``ordinary'' collinear modes we look at next, but still has a degree of collinearity in its angular scaling, due to measurement of small $e_2^{(\alpha)}$. It can resolve the phase space boundary around the region labeled $S_C$ in \fig{ztheta}, but none others.
This mode may or may not pass soft-drop and contributes to the measurement only when it passes soft-drop. This mode is then identical to the one in the case of light quark jets \cite{Frye:2016aiz} and is insensitive to the mass of the heavy quark. The collinear-soft scale $\mu_{cs} \sim E_J (e_2^{(\alpha)})^{1/\alpha} z_\text{cut}^{1-\frac{1}{\alpha}}$, because $z_\text{cut}<1$ and $\alpha<1$, is actually larger than the ``ordinary'' collinear scale we will consider below. Thus, at this point, we can then further factorize the cross section as in \cite{Frye:2016aiz}, 
\be
\label{eq:factorization2}
\frac{d \sigma}{d e_2^{(\alpha)}} = \sigma_0(Q^2,R,\mu) \times S\left(E_J z_\text{cut},\mu\right) \times S_c\left( E_J z_\text{cut}(e_2^{(\alpha)}/z_\text{cut})^{1/\alpha},\mu\right) \otimes J_q\left( e_2^{(\alpha)}, m_q/E_J,\mu\right), 
\ee
where as explained in \cite{Frye:2016aiz}, the collinear-soft function $S_c$, depends only on the single scale shown, related to its virtuality. Dependence on all lower, more collinear, scales is still in the function $J_q$, which we need to factorize further. (The $\otimes$ in \eq{factorization2} indicates a convolution in the variable $e_2^{(\alpha)}$.)

\paragraph{Collinear and ultracollinear modes} 

In \eq{factorization2}, $J_q$ is the massive quark jet function, defined by the matrix element 
\bea
\label{eq:jet}
J_q(e_2^{(\alpha)}) =\frac{(2\pi)^3}{N_C} \Tr \langle 0|\frac{\slashed{\bar n}}{2}\chi_n(0)\delta(\omega-\bar n\cdot \mathcal{P})\delta^{(2)}(\vec{\mathcal{P}})\delta(e_2^{(\alpha)}-\hat e_2^{(\alpha)})\bar \chi_n(0)|0\rangle
\eea
where $n$ is the light-like direction of the jet and $\omega= 2 E_J$, where $E_J$ is the energy of the jet. The field $\chi_n$ represents a quark with mass $m_q$ moving in the $n$ direction \cite{Leibovich:2003jd}. The $\mathcal{P^\mu}$ operator fixes the large part (``label'') of the collinear momentum to be $\omega n^\mu/2$ \cite{Bauer:2001ct}. The measurement operator $\hat e^{(\alpha)}_2$ is defined by its action on a collinear state $\ket{X_n}$, taking the collinear limit of the full measurement function:
\bea
\label{eq:e2hat}
\hat e_2^{(\alpha)}\ket{X_n} = \frac{2^{3\alpha/2}}{\omega^2} \sum_{i<j \in X_n}(\bar n\cdot p_i)^{1-\alpha/2} (\bar n \cdot p_j)^{1-\alpha/2}(p_i \cdot p_j)^{\alpha/2}\ket{X_n}
\eea
The jet function \eq{jet}  contains all the modes at and below this scale that are potentially sensitive to soft drop. 

Let us consider the modes that make up this jet function. The relevant modes depend on how the angle of the collinear radiation $\theta_c$ from the initiating heavy quark compares to $\theta_\text{min}$ in \eq{thetamin}. There are two cases:
\begin{itemize}
\item $\theta_c > \theta_\text{min}$ \\
In this case, the scaling of the angle for this mode is set by the measurement and not by the mass of the quark. Its energy fraction is not limited, so $z_c\sim 1$. This corresponds to measurement values $e_2^{(\alpha)} \geq (m_q/E_J)^{\alpha}$. The light-cone momenta of the mode $(p_c)$ then scale as
\bea
p_c \sim E_J\left(1, \left(e_2^{(\alpha)}\right)^{2/\alpha},  \left(e_2^{(\alpha)}\right)^{1/\alpha}\right), 
\eea
which is insensitive to the quark mass. This region actually exists as a separate EFT region only for $\alpha>1$, illustrated in \fig{ztheta2}, in the lower region labeled $J_q$. Otherwise values of $e_2^{(\alpha)}$ in this regime are already in the nonsingular fixed-order region. The jet function thus becomes independent of the quark mass and is the same as  for a massless jet. In this regime, the factorized cross section then becomes 
\be
\label{eq:factorizationmassless}
\frac{d \sigma}{d e_2^{(\alpha)}} = \sigma_0(Q^2,R,\mu) \times S(E_J z_\text{cut},\mu) \times S_c\left( E_J z_\text{cut}(e_2^{(\alpha)}/z_\text{cut})^{1/\alpha},\mu\right) \otimes J_q\left( E_J (e_2^{(\alpha)})^{1/\alpha},\mu\right), 
\ee
which is the same form obtained in \cite{Frye:2016aiz} for massless quarks.
\item  $\theta_c \sim \theta_\text{min}$ \\ 
Once the angle $\theta_c$ hits $\theta_\text{min}$, it can go no lower, and thus the angular scaling of collinear modes in this regime is fixed. For $z\sim 1$, it contributes $e_2^{(\alpha)}\sim (\theta_\text{min})^\alpha$. To go to any lower values of $e_2^{(\alpha)}$, the only way to do so is to lower the energy fraction of the emitted collinear radiation, down to the lower limit $e_{2,\text{min}}^{(\alpha)}$ in \eq{e2min}, moving along the right-most red line in \fig{ztheta} or \fig{ztheta2}. The measurement of such $e_2^{(\alpha)}$ then constrains that $z\sim z_{uc}$, where
\bea
z_{uc} \sim e_2^{(\alpha)}/ (\theta_{\min})^{\alpha} 
\eea
and the mode's momentum scales as 
\bea
\label{eq:puc}
p_{uc} &\sim & E_J  e_2^{(\alpha)}/ (\theta_\text{min})^{\alpha} \left( 1, \theta_\text{min}^2, \theta_\text{min} \right)\nn\\
&\equiv& m_q e_2^{(\alpha)}/ (\theta_\text{min})^{\alpha}\left( E_J/m_q, m_q/E_J, 1\right) \nn\\
&\equiv & \Gamma \left( v_+, v_-, 1 \right),
\eea
where we have defined, 
\bea
\label{gamma}
\Gamma =  m_q e_2^{(\alpha)}/ (\theta_{\min})^{\alpha},~~~~~~~~v^{\mu} \equiv \left( E_J/m_q, m_q/E_J,\vect{0}_\perp \right) \,.
\eea
This way (\eq{puc}) of writing this mode's momentum suggests that it is the momentum of a small fluctuation around a \emph{boosted} heavy quark at nonzero $e_2^{(\alpha)}$. Here $v^{\mu}$ is the four velocity of the boosted heavy quark, given again in light-cone coordinates. This mode is the ultra-collinear mode of boosted HQET (bHQET) \cite{Hoang:2017kmk,Fleming:2007xt}. This tells us that we need to match the massive jet function onto a boosted HQET jet function with the heavy quark mass $m_q$ (which is now like a hard scale) integrated out. The scale $\Gamma$ serves as the IR scale for this EFT. It is analogous to the top quark width which served as $\Gamma$ in \cite{Hoang:2017kmk,Fleming:2007xt}, but here is due simply to the radiation from the heavy quark. We can immediately see that $z_{uc} > z_\text{cut}$ so that, as expected, this mode automatically passes soft drop. In this region of $e_2^{(\alpha)}$, the cross section \eq{factorization2} factorizes further as
\begin{align}
\label{eq:factorizationbHQET}
 \frac{d \sigma}{d e_2^{(\alpha)}} &= \sigma_0(Q^2,R,\mu) \times S\left(E_J z_\text{cut},\mu\right) \times H(m_q,\mu) \\
 &\qquad \times S_c\left( E_J z_\text{cut}(e_2^{(\alpha)}/z_\text{cut})^{1/\alpha},\mu \right) \otimes B_+(\Gamma,\mu) \nn
\end{align}
where $H(m_{q})$ is a Wilson coefficient from integrating out $m_q$ to match onto bHQET, and $B_+$ is the bHQET jet function \cite{Hoang:2017kmk,Fleming:2007xt}.

\end{itemize}

Finally, we note that when $e_2^{(\alpha)}$ reaches $e_{2,\text{min}}^{(\alpha)}$ defined in \eq{e2min}, and as illustrated in \fig{ztheta} and \fig{ztheta2}, the collinear-soft and ultracollinear modes actually merge:
\begin{align}
\text{\eq{pcs}}\Rightarrow p_{cs} &\rightarrow z_\text{cut} E_J\left( 1, \theta_\text{min}^2,\theta_\text{min}\right) \\
\text{\eq{puc}}\Rightarrow p_{uc} &\rightarrow z_\text{cut} E_J \left( 1, \theta_\text{min}^2,\theta_\text{min}\right)\,.
\end{align}
We should then match to a new function $B_+^{SD}$, which is simply the bHQET jet function with an explicit soft drop constraint, with the Wilson coefficient $H(m,\mu)$ still factorized out. This entails a computation without expanding in the scale hierarchy $e_2^{(\alpha)}/e_{2,\text{min}}^{(\alpha)}$, which is no longer small in this region. In this region of $e_2^{(\alpha)}$, the cross section \eq{factorization2} factorizes as
\begin{align}
\label{eq:factorizationmin}
 \frac{d \sigma}{d e_2^{(\alpha)}} &= \sigma_0(Q^2,R,\mu) \times S\left(E_J z_\text{cut},\mu\right) \times H(m_q,\mu)\times B^{SD}_+(\Gamma,\mu, z_{\text{cut}}) \,.
\end{align}
We compute each of these functions to one loop order in \sec{oneloopEFT}. There is also a fixed-order matching required between the leading-order combination $S_c\otimes B_+$ in \eq{factorizationbHQET} to $B_+^{SD}$ in \eq{factorizationmin} to capture terms power-suppressed in $e_2^{(\alpha)}/e_{2,\text{min}}^{(\alpha)}$ in the former region but not the latter. We perform this matching, which is a novel feature arising for shape observables with a nonzero minimum, in \sec{resum}.


\subsection{Regions of EFTs}
\label{ssec:EFTregions}

\begin{figure}
\centerline{\scalebox{.45}{\includegraphics{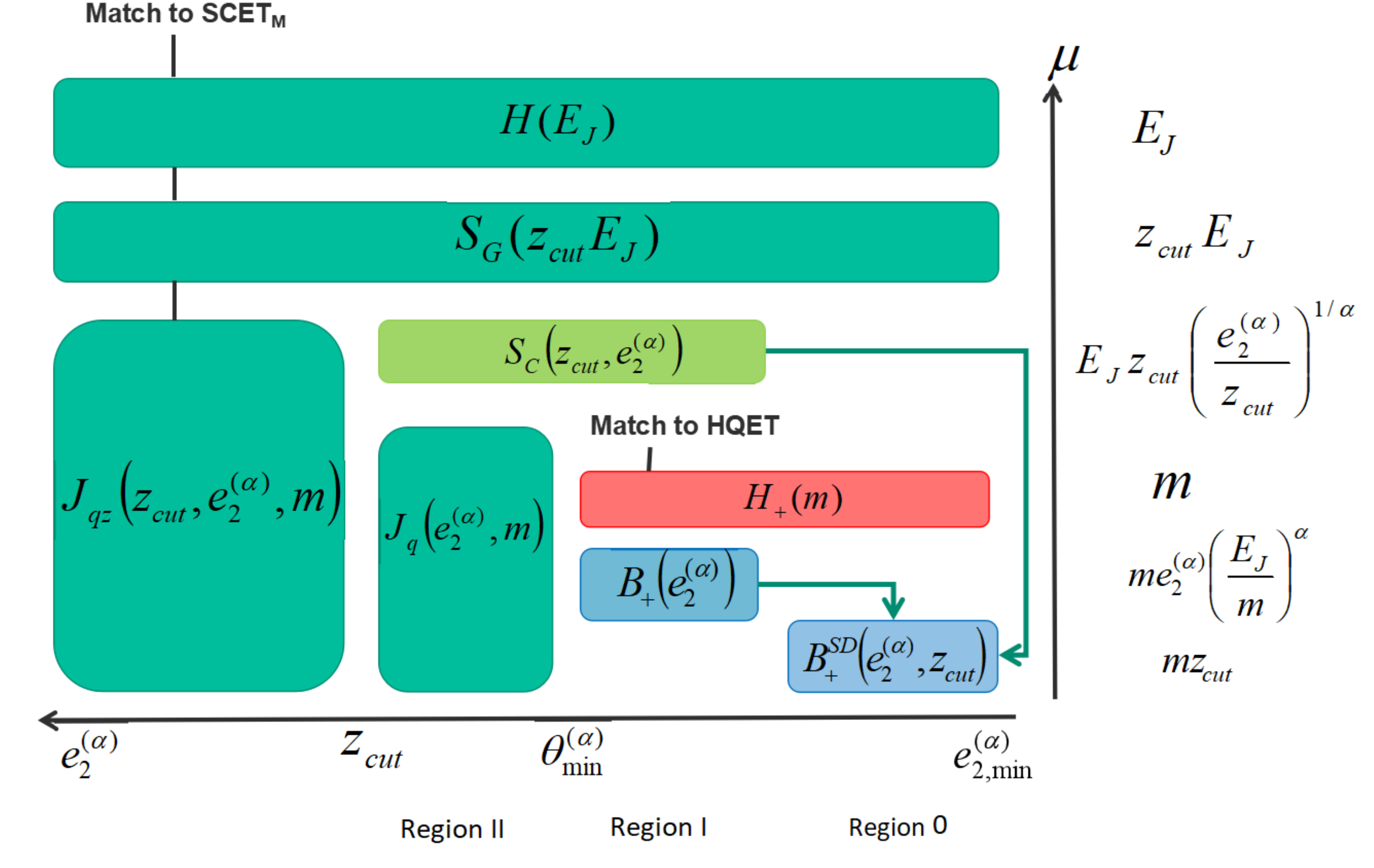}}}
\vskip-0.5cm
\caption[1]{Hierarchy of scales in factorization. The horizontal axis indicates the size of $e_2^{(\alpha)}$ (increasing right to left) while the vertical axis shows increasing virtuality of the relevant modes in each region. See \ssec{EFTregions} for descriptions of the regions and the applicable factorizations.}
\label{fig:scales} 
\end{figure}

\begin{figure}
\centerline{\scalebox{.45}{\includegraphics{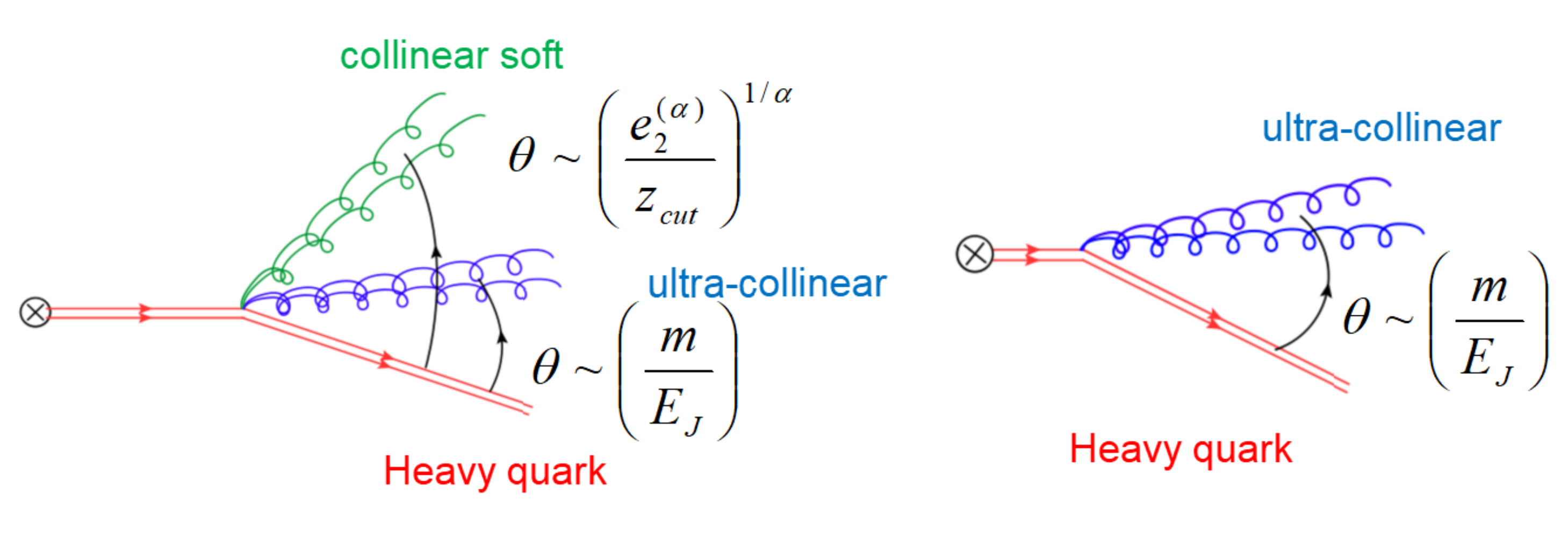}}}
\vskip-0.5cm
\caption[1]{Radiation modes for regions I and 0 of factorization. In region I (left), there are two radiation modes that contribute, the ultra-collinear and collinear-soft mode. In region 0 (right), these modes merge into a single soft drop constrained ultra-collinear mode.}
\label{fig:rad} 
\end{figure}

Hence  the applicable EFT now depends on the value of $e_2^{(\alpha)}$ relative to the scale $(m_q/E_J)^\alpha$, at the threshold where the ``ordinary'' collinear modes above hit the angle $\theta_\text{min}$ and, below this scale, have to be matched onto bHQET ultracollinear modes. 
\begin{itemize}

\item \emph{Region I:} $ e_{2,\text{min}}^{(\alpha)} < e_2^{(\alpha)}  < (m_q/E_J)^{\alpha} $\\
In this range,  radiation collinear to the heavy quark is restricted to have an angle $\sim\theta_\text{min}$ (see \fig{rad}), and thus the $e_2^{(\alpha)}$ cross section is sensitive to the mass of the quark. The applicable factorization is \eq{factorizationbHQET}.

\item \emph{Region II:} $ (m_q/E_J)^{\alpha} < e_2^{(\alpha)} < z_\text{cut}$ \\
In this range, radiation collinear to the heavy quark has an angle larger than $\theta_\text{min}$, and is instead restricted by the measurement of $e_2^{(\alpha)}$ itself. Then the cross section is insensitive to the mass and essentially behaves as for a massless groomed jet, and we have the factorization \eq{factorizationmassless}.

\item \emph{``Region 0'':} $ e_2^{(\alpha)} \sim e_{2,\text{min}}^{(\alpha)}$ \\
In this range,  radiation collinear to the heavy quark is restricted to have an angle $\sim\theta_\text{min}$. At the same time it is sensitive to the grooming parameter $z_\text{cut}$. There is no independent collinear-soft mode, as it merges with the ultracollinear mode (see \fig{rad}), and we have the factorization \eq{factorizationmin}.
\end{itemize}
These regions are indicated by the bottom right columns in \fig{scales}.

A natural question is whether the transition between regimes happens in a smooth manner, especially at the boundary $e_2^{(\alpha)} = (m_q/E_J)^{\alpha}$. 
First of all, we observe that at the value $e_2^{(\alpha)} =  (m_q/E_J)^{\alpha}$, the ultra-collinear and collinear mode have the same scaling 
\bea
p_c \sim  E_J \left( 1, \frac{m_q^2}{E_J^2}, \frac{m_q}{E_J} \right) \,.
\eea
At the same time, not surprisingly, the massive and massless jet function defined by \eq{jet} will turn out to be identical at this value of $e_2^{(\alpha)}$, as we will see in \ssec{jet}. These properties ensure that the distribution in $e_2^{(\alpha)}$ is indeed continuous at the transition point, as we will see explicitly in \sec{resum}.

Now, it is possible, depending on the relative sizes of $z_\text{cut}$ and $(m_q/E_J)^\alpha$, that the region II EFT simply doesn't exist. This will happen if $(m_q/E_J)^\alpha\gtrsim z_\text{cut}.$ For typical values of these parameters, $m_q\sim 5$ GeV, $E_J\sim 100$ GeV, and $z_\text{cut}\sim 0.1$, this can only happen if $\alpha<1$. And indeed, we choose to consider $\alpha\sim 0.5$ in what follows. Then the scales satisfy $(m_q/E_J)^\alpha \sim 0.2 \gtrsim z_\text{cut}.$
In such a case, only the region I EFT above exists. This is the situation illustrated in \fig{ztheta}. When $e_2^{(\alpha)}$ reaches the upper limit of the region, we simply match our massive EFT onto the full theory fixed-order cross section at the common scale $ (m_q/E_J)^{\alpha}$, without an explicit transition into the massless EFT regime. 

In the case that $(m_q/E_J)^\alpha\ll z_\text{cut}$, we would first transition from the massive (region I) to the massless (region II) EFT at $e_2^{(\alpha)} \sim (m_q/E_J)^{\alpha}$ and then match the massless EFT onto the full theory fixed-order cross section at the scale $z_\text{cut}$. This is illustrated in \fig{ztheta2}, where the value of $e_2^{(\alpha)}$ determines whether $J_q$ gets factored further into $S_C$ and $B_+$.\footnote{When region II exists, there may be a region near the boundary with region I where the virtuality of the SCET$_+$ modes is similar to $m_b$, and a further factorization with ``mass modes'' is required starting at 2 loops.}

\fig{ztheta} and \fig{ztheta2} illustrate clearly that the value of the angular exponent $\alpha$ in the definition of the correlator $e_2^{(\alpha)}$ determines whether or not one passes through a region II EFT before transitioning to a region I EFT. 
For us, the more interesting scenario is actually the first, in \fig{ztheta}, with only a region I EFT, since it means that the $e_2^{(\alpha)}$ distribution is sensitive to the mass over most of its range. 
So henceforth we will work with $\alpha <1$, and thus only have a region I EFT setup. We also note that this choice of $\alpha$ is what made the virtuality of the collinear-soft mode above to be greater than that of the HQET ultra-collinear mode, which is reflected in \fig{scales} and in our discussion of factorization above. We will see later that this has important consequences for the nonperturbative corrections to the $e_2^{(\alpha)}$ distribution.

\section{One-loop EFT results}
\label{sec:oneloopEFT}

In this section we compute to one-loop order the functions in the factorized cross section \eq{factorization2}.
All results are computed using dimensional regularization in the $\MSbar$ scheme.

\subsection{Massive quark jet function}
\label{ssec:jet}

The quark jet function is defined by the matrix element given in \eqs{jet}{e2hat}.
We use the Feynman rules for massive SCET \cite{Leibovich:2003jd}.

\begin{figure}
\centerline{\scalebox{.45}{\includegraphics{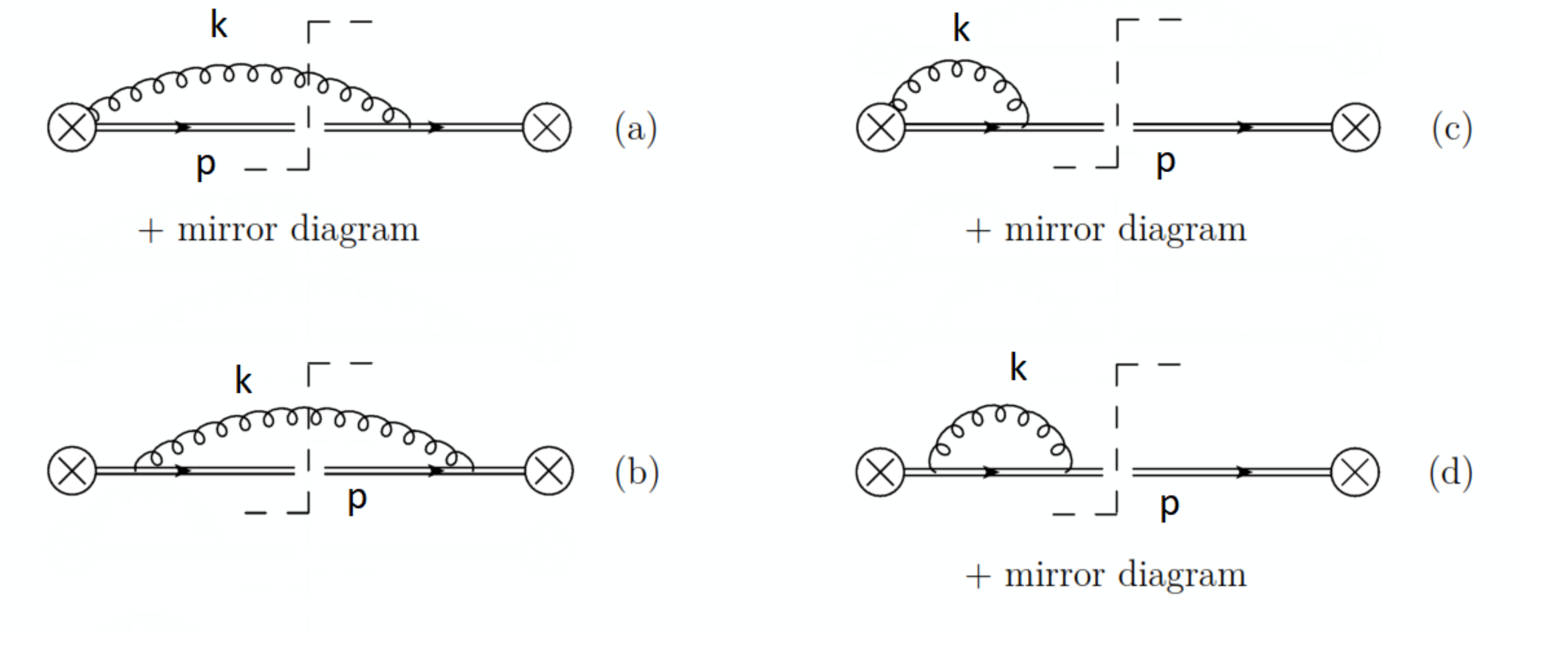}}}
\vskip-0.5cm
\caption[1]{Real and virtual diagrams contributing to the jet function.}
\label{fig:diag} 
\end{figure}

We have contributions from two real diagrams shown in \fig{diag}:
\bea
\label{eq:Ra}
R_a &=& 8 g^2C_F \tilde \mu^{2\epsilon} \int d^4 p\, \delta^+(p^2-m^2) \int \frac{d^dk}{(2\pi)^{d-1}}\delta(\omega -p^--k^-) \delta^+(k^2) \delta^2( k_{\perp}+p_{\perp})\\
&& \times\frac{p^-(p^-+k^-)}{k^- ( (p+k)^2-m^2)}\delta \left(e_2^{(\alpha)}- e_2^{(\alpha)}(p,k)\right) \nn
\eea
and
\bea
\label{eq:Rb}
R_b &=&  -4 g^2 C_F  \tilde \mu^{2\epsilon} \int d^4 p\, \delta^+(p^2-m^2) \int \frac{d^dk}{(2\pi)^{d-1}}\delta(\omega -p^--k^-) \delta(k^2) \delta^2( k_{\perp}+p_{\perp}) \\
&&\times \frac{p^-(k^-+p^-)^2}{[(p+k)^2-m^2]^2} \left (\frac{4m^2}{p^-(p^-+k^-)}-\frac{p_{\perp}^2+m_q^2}{(p^-)^2}-\frac{m^2}{(p^-+k^-)^2}\right)\delta(e_2^{(\alpha)}- e_2^{(\alpha)}(p,k)) \,, \nn
\eea
where 
\bea
e_2^{(\alpha)}(p,k)  = \frac{p^- k^-}{\omega^2} \left( 4\frac{(p+k)^2-m_q^2}{p^- k^-}\right)^{\alpha/2} \,,
\eea
and the scale $\tilde\mu^2 = \mu^2 e^{\gamma_E}/(4\pi)$ per the $\MSbar$ scheme.
Here and below we define $\delta^+(p^2) \equiv \delta(p^2) \theta(p^0)$ for a four-vector $p$.

It is easier to compute these integrals in Laplace space, in which the jet function is defined
\be
\widetilde J(s,\mu) = \int_0^\infty de_2^{(\alpha)} e^{-se_2^{(\alpha)}} J(e_2^{(\alpha)},\mu)\,,
\ee
for which the renormalized (finite) result in the $\MSbar$ scheme for the diagrams $R_{a,b}$ yields
\bea
\widetilde R_{a}(\mu) &=& -\frac{\alpha_s}{\pi}  \frac{C_F}{ (1-\alpha)}  \ln^2\left(\frac{e^{\gamma_E} s\mu (4\Delta)^{(\alpha-1)/2}}{\omega}\right)\\
\widetilde R_{b}(\mu) &=&  \frac{\alpha_s}{\pi}C_F \ln\left(\frac{e^{\gamma_E} s\mu (4\Delta)^{(\alpha-1)/2}}{\omega}\right) \nn
\eea
where $\Delta \equiv m_q^2/\omega^2$. 

There are two contributions from virtual diagrams (\fig{diag}):
\be
V_a =  4 iC_F g^2  \tilde \mu^{2\epsilon} \int \frac{d^dk}{(2\pi)^d} \frac{\omega-k^-}{[(p-k)^2-m^2+i0](k^--i0)(k^2+i0)} ,
\ee
whose finite result is
\be
V_a(\mu)  = \frac{\alpha_s C_F}{\pi} \left( \ln^2\left(\frac{\mu}{m} \right)+2 \ln \left(\frac{\mu}{m} \right) \right)\,,
\ee
and
\bea
 V_b &=& - 4i C_F g^2  \tilde \mu^{2\epsilon} \int d^4p \frac{\delta^+(p^2-m^2-\Lambda^2)}{\Lambda^2} \delta^2(\vec{p}_{\perp}) \delta(\omega-p^-) (p^-)^2\\
&\times& \int \frac{d^d k}{(2\pi)^d} \frac{1}{ (k^2+i0) [(k+p)^2-m^2+i0] }\left(\frac{4m^2}{p^-}-(p^++k^+)-\frac{m^2(p^-+k^-)}{(p^-)^2}\right)\nn
\eea
with $\Lambda^2$ being the off-shellness of the quark. To extract the residue of the propagator, we Taylor expand the integrand in $\Lambda^2$ and retain the $\Lambda^0$ term which gives us the renormalized(finite) result,
\bea
 V_{b}(\mu) &=&-3 \frac{\alpha_s}{2\pi} C_F  \ln \left(\frac{\mu}{m_q}\right).
\eea
\\
Combining the results for real and virtual emissions together with the tree-level result, we get the renormalized result
\bea
\label{eq:massivejet}
\widetilde J(s, \mu)=  1+ \frac{\alpha_s C_F}{\pi}\left( \frac{1}{\alpha-1} L_C^2 +L_V^2 + L_C +\frac{1}{2}L_V + \frac{\pi^2}{8(\alpha-1)}- \frac{\pi^2}{24} +\frac{\alpha}{2}+1\right)
\eea
with 
\bea
\label{eq:LCLV}
L_C &=& \ln \left(\frac{\mu s e^{\gamma_E}}{\omega} (4\Delta)^{\frac{\alpha-1}{2}}\right)  \nn\\
L_V &=& \ln \left(\frac{\mu}{m_q}\right).
\eea
We have checked that the result \eq{massivejet} when $\alpha=2$ agrees with the one-loop massive quark jet function for jet mass derived in \cite{Fleming:2007xt}.

These results give us the anomalous dimension for the jet function,
\bea
\gamma_{\mu}^J = \frac{\alpha_s C_F}{\pi} \left(\frac{\alpha}{\alpha-1} \ln \left(\frac{\mu^2 (s e^{\gamma_E})^{2/\alpha}}{\omega^2}\right)+\frac{3}{2} \right).
\eea
Unsurprisingly, since the global soft, collinear-soft and the hard function remain unchanged compared to the massless groomed jet, the anomalous dimension is the same as that of the massless jet function. 
Comparing with \cite{Frye:2016aiz}, we also observe that the massive and the massless jet function have identical values at $1/s \sim (m/E_J)^{\alpha}$, which marks the transition point between the massive and massless regime.

\subsection{Global Soft}
\label{GSoft}
The global soft function is defined as the following matrix element of Wilson lines 
\bea
S_G(E_Jz_\text{cut}) = \frac{1}{2N_C} \text{Tr} \langle 0|T\{Y_nY_{\bar n}\}\widehat \Theta_{SD} \widehat\Theta_R\bar T\{Y_n^\dag Y_{\bar n}^\dag\}|0\rangle \,.
\eea
$\widehat \Theta_{SD}$ denotes the soft drop groomer. $\widehat\Theta_R$ imposes the jet radius constraint. We require that the global soft modes fail soft drop.  From the literature \cite{Frye:2016aiz,Ellis:2010rwa,Chien:2015cka}, we can write down the result for the one loop renormalized function  
\bea
S_G(E_Jz_\text{cut},\mu) = 1+\frac{\alpha_s C_F}{\pi}\left( \ln^2 \left(\frac{\mu}{2E_J z_\text{cut}}\right)-\frac{\pi^2}{8} \right) \,,
\eea
so the natural scale for this function is $\mu_{gs} = 2E_J z_\text{cut}$.\footnote{For a finite $R$, the scale would be modified to $\mu_{gs} = 2E_J z_{\text{cut}}R$. However, since we are working in a regime $R\sim 1$, we set $R=1$ for this calculation, i.e. we choose not to resum any logarithms in $R$ since they are small, essentially giving the result for a hemisphere jet. The $R$ dependence becomes important in the tail region where we match onto the full theory fixed-order cross section and is implemented numerically.}
This leads to the anomalous dimension 
\bea
\gamma_{\mu}^{SG} =  \frac{\alpha_s C_F}{\pi}\ln\left(\frac{\mu^2}{4E_J^2 z^2_\text{cut}}\right) \,.
\eea

\subsection{Collinear-soft function}
\label{CS}
As explained in Section \ref{modes}, the collinear-soft function is not affected by the mass of the heavy quark. Hence, it remains exactly the same as that for a massless jet. We reproduce the result (\cite{Frye:2016aiz}) for completeness. The collinear-soft function is defined as the matrix element
\bea 
S_c(z_\text{cut},e_2^{(\alpha)}) = \frac{1}{N_C} \Tr \langle 0|\text{T} \left(Y_n^{\dagger} W_t\right) \delta \left(e_2^{(\alpha)} - \left(1-\hat \Theta_{SD}\right)\hat e_2^{(\alpha)} \right)\bar T\left(W_t^{\dagger} Y_n\right)|0 \rangle, 
\eea
The $W_t$ Wilson line is the same one that appears in the massive quark jet function but is composed of collinear-soft fields so that 
\bea
W_t = P \exp\biggl[\int_{-\infty}^0 ds\, \bar n \cdot A_{cs}^a(x+s\bar n) T^a \biggr]
\eea 
where we have defined 
\bea
 e_2^{(\alpha)}|X_{S_c} \rangle = \frac{2^{\alpha}}{2 \omega} \sum_{i \in X_{S_c}} (\bar n \cdot p_i)^{1-\alpha/2}(n \cdot p_i)^{\alpha/2}|X_{S_c} \rangle.
\label{Cmeasure} 
\eea
The collinear-soft modes only contribute to the measurement if they pass soft drop, which is implemented by the $\hat \Theta_{SD}$ term. 
In Laplace space, we have the result for the renormalized (finite) function is
\bea
\widetilde S_{c}(z_\text{cut},s,\mu) = 1+\frac{\alpha_s C_F}{2\pi} \left[-\frac{2\alpha}{(\alpha-1)}L^2_{S_c}+\frac{\pi^2}{12}\frac{(\alpha+2)(\alpha-2)}{\alpha(\alpha-1)}\right]
\eea
where 
\bea
L_{S_c} = \ln \frac{\mu (s e^{\gamma_E} )^{1/\alpha}}{E_J (z_\text{cut})^{\frac{\alpha-1}{\alpha}}}
\eea
and this leads to the anomalous dimension 
\bea
 \gamma_{\mu}^{S_c} = -C_F \frac{\alpha_s}{\pi}\frac{\alpha}{\alpha-1} L_{S_c}.
\eea


\subsection{Boosted HQET jet function}
We evaluate the boosted HQET jet function at one loop in two regimes, one in which the ultra-collinear mode automatically passes soft drop, which happens when $e_2^{(\alpha)}\gg e_{2,\text{min}}^{(\alpha)}$ and a second region where we explicitly impose the soft-drop constraint in the region $e_2^{(\alpha)} \sim e_{2,\text{min}}^{(\alpha)}$ (the upper-right corner region of \fig{ztheta} or \fig{ztheta2}).

\subsubsection{$e_2^{(\alpha)}\gg e_{2,\text{min}}^{(\alpha)}$}
We define the jet function in boosted HQET as,
\bea
B_+ = \frac{1}{\mathcal{N}} \langle 0 | \bar h_{v_+} W_n \delta(e_2^{(\alpha)}-\hat e_2^{(\alpha)}) W_n^{\dagger} h_{v_+} |0 \rangle . 
\eea
$v_+$ is the velocity of the boosted heavy quark $ v_+ = ( m/\omega, \omega/m, \vect{0}_{\perp})$ and the residual momentum that make up the modes of this jet function (known as ultracollinear modes) scales as 
\bea
k^{\mu} \sim \Gamma \left(m/\omega, \omega/m, \vect{1}_\perp \right)  
\eea 
where  $\Gamma = m e_2^{(\alpha)}/(\Delta)^{\alpha/2}$ is the IR scale for this EFT ($m$ being the hard scale).

We can also write down the result of the measurement function $\hat e_2^{(\alpha)}$ acting on a state with one gluon of momentum $k$ emitted from the heavy quark as,
\bea
\label{Hmeasure}
e_2^{(\alpha)}(k) = \frac{\bar n \cdot k}{\omega} \left( 4\frac{ m v_+ \cdot k}{\omega \bar n \cdot k} \right)^{\alpha/2}. 
\eea
At one loop, the diagrams again are the same as in \fig{diag}. We have two contributions:
\bea
R_1&=& 2g^2 C_F \tilde \mu^{2\epsilon} \int \frac{d^dk}{(2\pi)^d}\delta^+(k^2) \frac{\bar n \cdot v_+}{[v_+ \cdot k] \bar n \cdot k} \delta (e_2^{(\alpha)} -e_2^{(\alpha)}(k))\nn\\
&=&\frac{\alpha_s C_F }{\pi \Gamma[1-\epsilon](e_2^{(\alpha)})^{1+2\epsilon}}\left(\frac{\tilde \mu (4\Delta)^{\alpha/2}}{m }\right)^{2\epsilon}\left(\frac{1}{-\epsilon(\alpha-1)}+\epsilon \frac{\pi^2}{6}\right)\,,
\eea
which is the bare result and is exactly the same as the integral $R_a$ evaluated in the previous section.
We have one more diagram which evaluates to 
\bea 
R_2 =  g^2C_F \int d^d k \delta(k^2) \frac{ (v_+)^2}{(v_+ \cdot k)^2}\delta(e_2^{(\alpha)} - e_2^{(\alpha)}(k))
\eea
which turns out to be the same as $R_b$ in \eq{Rb}.

The virtual diagrams are all scaleless and disappear in dimensional regularization. The one loop renormalized result of the HQET jet function in Laplace space then yields 
\bea
\label{eq:bHQETjet}
\widetilde B_{+}^{(1)}(s,\Delta,\mu) =  \frac{\alpha_s C_F}{\pi}\left( \frac{1}{\alpha-1} L_C^2 + L_C + \frac{\pi^2}{8(\alpha-1)}- \frac{\pi^2}{12} +\frac{\alpha}{2}\right). 
\eea 
The matching between $\tilde J$ in \eq{massivejet} and $\widetilde B_+$ in \eq{bHQETjet} then tells us that the matching function $H$ in \eq{factorizationbHQET} is
\bea
H_{+}(m, \mu)=  1+ \frac{\alpha_s C_F}{\pi}\left( L_V^2 +\frac{1}{2}L_V +\frac{\pi^2}{24}+1\right) 
\eea
i.e., all the virtual corrections of the SCET jet function. This also makes sense since it gives us a clean separation of the scales $m$ and $ m e_2^{(\alpha)}/ (4\Delta)^{\alpha/2}$.

The anomalous dimensions for these functions are now given as 
\bea
\gamma_{\mu}^{B_+} &=&   \frac{\alpha_s C_F}{\pi ({\alpha-1})} \ln \frac{\mu^2 s^2e^{2\gamma_E} (4\Delta)^{\alpha-1}}{\omega^2} + \frac{\alpha_s C_F}{\pi}\nn\\
\gamma_{\mu}^{H_+} &=& \frac{\alpha_s C_F}{\pi} \ln \frac{\mu^2}{m^2}+\frac{\alpha_s C_F}{2\pi} \,.
\eea
The matching onto bHQET agrees with that obtained in \cite{Fleming:2007xt} (accounting for us having one heavy jet instead of two). In fact it is possible to obtain all singular logarithmic terms for the observable which is sensitive to the heavy quark mass simply from knowing the collinear-soft function and the matching coefficient from massive SCET to boosted HQET (which is independent of the specific observable considered). So an easy way to extend our results to NNLL accuracy would be the computation of the two-loop collinear-soft function.

\subsubsection{$e_2^{(\alpha)} \sim e_{2,\text{min}}^{(\alpha)}$}

We define the jet function in boosted HQET with an explicit soft drop constraint as,
\bea
B_+^{SD} = \frac{1}{\mathcal{N}} \langle 0 | \bar h_{v_+} W_n  \delta \left(e_2^{(\alpha)} - \left(1-\hat \Theta_{SD}\right)\hat e_2^{(\alpha)} \right) W_n^{\dagger} h_{v_+} |0 \rangle. 
\eea
There are two diagrams as before, in \fig{diag}, and for each of them, we can explicitly divide up the phase space in terms of regions that pass or fail soft drop.
\bea
R_{1} &=& 2g^2 C_F \tilde \mu^{2\epsilon} \int \frac{d^dk}{(2\pi)^d}\delta^+(k^2) \frac{\bar n \cdot v_+}{[v_+ \cdot k] \bar n \cdot k} \delta (e_2^{(\alpha)} -e_2^{(\alpha)}(k))\theta(\bar n \cdot k - \omega z_\text{cut})\nn\\  
&+&2g^2 C_F \tilde \mu^{2\epsilon}  \delta (e_2^{(\alpha)})\int \frac{d^dk}{(2\pi)^d}\delta^+(k^2) \frac{\bar n \cdot v_+}{[v_+ \cdot k] \bar n \cdot k}\theta(\omega z_\text{cut}-\bar n \cdot k )\nn\\  
&\equiv& R_{1,a}+R_{1,b} \,.
\eea
Due to the phase space and measurement constraints, $R_{1,a}$ is finite and can be directly evaluated in 4 dimensions. $R_{1,b}$ contains a divergence. The renormalized results for these diagrams give us 
\bea
 R_{1,a}(\mu) &=& \frac{\as C_F}{\pi}\theta\bigl(e_2^{(\alpha)} - e_{2,\text{min}}^{(\alpha)}\bigr) \frac{2}{\alpha} \frac{1}{e_2^{(\alpha)} }\ln \frac{e_2^{(\alpha)}}{e_{2,\text{min}}^{(\alpha)}}\nn\\
R_{1,b}(\mu) &=& - \delta (e_2^{(\alpha)})\frac{\alpha_s C_F}{\pi}\left(\ln^2\left( \frac{\mu}{mz_\text{cut}}\right)+\frac{\pi^2}{24}\right) \,.
\eea

We have one more diagram which can likewise be split up into two pieces. 
\bea
R_2 &=&  g^2 \tilde\mu^{2\epsilon}C_F \int d^d k \delta^+(k^2) \frac{ (v_+)^2}{(v_+ \cdot k)^2}\delta(e_2^{(\alpha)} - e_2^{(\alpha)}(k))\theta(\bar n \cdot k - \omega z_\text{cut})\nn\\
&+&   g^2 \tilde\mu^{2\epsilon} C_F\delta(e_2^{(\alpha)}) \int d^d k \delta^+(k^2) \frac{ (v_+)^2}{(v_+ \cdot k)^2}\theta( \omega z_\text{cut}-\bar n \cdot k ) \nn\\
&=& R_{2,a}+R_{2,b}  \,.
\eea
The renormalized (finite) result is then
\bea
R_{2,a}(\mu) &=& \frac{\alpha_sC_F}{\pi} \theta\bigl(e_2^{(\alpha)} - e_{2,\text{min}}^{(\alpha)}\bigr) \frac{1}{e_2^{(\alpha)}}\left(1-\left(\frac{e^{(\alpha)}_{2,\text{min}}}{e_2^{(\alpha)}}\right)^{2/\alpha}\right)\nn\\
R_{2,b}(\mu) &=& \delta (e_2^{(\alpha)})\frac{\alpha_sC_F}{\pi}\ln\left( \frac{\mu}{mz_\text{cut}}\right) \,.
\eea
Putting all the pieces together gives us 
\bea
 B_{+}^{SD(1)}(e_2^{(\alpha)},m,\mu) &=&  \frac{\alpha_s C_F}{\pi} \theta\bigl(e_2^{(\alpha)} - e_{2,\text{min}}^{(\alpha)}\bigr)\left( \frac{2}{\alpha} \frac{1}{e_2^{(\alpha)} }\ln \frac{e_2^{(\alpha)}}{e_{2,\text{min}}^{(\alpha)}}+ \frac{1}{e_2^{(\alpha)}}\left(1-\left(\frac{e^{(\alpha)}_{2,\text{min}}}{e_2^{(\alpha)}}\right)^{2/\alpha}\right)\right)\nn\\
&+&  \delta (e_2^{(\alpha)})\frac{\alpha_s C_F}{\pi}\left(-\ln^2\left( \frac{\mu}{mz_\text{cut}}\right)+\ln\left( \frac{\mu}{mz_\text{cut}}\right)-\frac{\pi^2}{24}\right) \,.
\label{eq:bjetSD}
\eea
The result at nonzero $e_2^{(\alpha)}$ reproduces the one-loop fixed-order result we will compute later in the singular limit \eq{collinearFO}. 
The anomalous dimension contribution (in Laplace space) is entirely from the piece that fails soft drop,
\bea
 \gamma_{\mu}^{B_+^{SD}} = \frac{\alpha_s C_F}{\pi} \left( -2\ln\left( \frac{\mu}{mz_\text{cut}}\right)+1 \right)
\eea
which we can verify is the combined anomalous dimension of the unconstrained HQET jet function $B_+$ and the collinear-soft function $S_c$ as desired by RG invariance.
\bea
 \gamma_{\mu}^{B_+^{SD}} = \gamma_{\mu}^{B_+}+ \gamma_{\mu}^{S_c}\,.
\eea

\subsection{Consistency of RG equations}
At each step of factorization we can verify the consistency of RG equations 
\bea
\gamma_{\mu}^J+ \gamma_{\mu}^{S_c}+ \gamma_{\mu}^{S_G}+\gamma_{\mu}^H &=& 0\nn\\
\gamma_{\mu}^{B_+}+\gamma_{\mu}^{H_+}+\gamma_{\mu}^{S_c}+ \gamma_{\mu}^{S_G}+\gamma_{\mu}^H &=& 0\nn\\
\gamma_{\mu}^{B_+^{SD}}+\gamma_{\mu}^{H_+}+ \gamma_{\mu}^{S_G}+\gamma_{\mu}^H &=& 0\,,
\label{RG:consistency}
\eea
where the hard anomalous dimension is given as \cite{Bauer:2011uc,Ellis:2010rwa}
\bea
 \gamma_{\mu}^{H} = \frac{\alpha_s C_F}{\pi} \left( 2\ln\left( \frac{\mu}{2E_J}\right)-\frac{3}{2} \right)\,.
\eea

\section{Fixed-order result}
\label{sec:fixed}

\subsection{Full range of $e_2^{(\alpha)}$}

To predict the distribution  over the full range of $e_2^{(\alpha)} $, we will need to match the resummed result to a fixed-order cross section. The massive EFT is valid in the region $e_{2,\text{min}}^{(\alpha)} < e_2^{(\alpha)} <  (m_b/E_J)^{\alpha}$ (``region I'') while the massless EFT inhabits the regime $ (m_b/E_J)^{\alpha}< e_2^{(\alpha)} < z_{\text{cut}}$ (``region II''). (Though we recall that in this paper we choose to work in cases where $\alpha<1$ and $(m_b/E_J)^\alpha \gtrsim z_\text{cut}$, so only the region I EFT exists.) As $e_2^{(\alpha)}$ nears $z_\text{cut}$, power corrections in $e_2^{(\alpha)}/z_\text{cut}$ become important and must be included to maintain the accuracy of the result across the complete range of $e_2^{(\alpha)}$. The usual procedure is to turn off the resummation and match the resummed result to a fixed-order result using a profile function. This enables the distribution to make a smooth transition to the fixed-order result around $e_2^{(\alpha)} \sim z_\text{cut}$. 

We now compute the fixed-order cross section at $\cO(\alpha_s)$.  We implement a $k_t$-type jet algorithm to isolate two jets (back-to-back) of radius $R \sim  1$. At the same time, we implement the soft drop jet grooming algorithm described in \ssec{soft-drop} to remove soft radiation from the jet. The grooming parameter is $z_\text{cut}$ which is the fraction of the energy of the hard scale ($E_J$). Any radiation with $E< z_\text{cut}E_J$ is removed from the jet and does not contribute to the measurement of $e_2^{(\alpha)}$.

The cross section for this process at one loop, with three final-state particles, is given by the formula:
\begin{align}
\label{eq:fixedordercs}
\frac{d\sigma}{de_2^{(\alpha)}} = \frac{1}{2Q^2} \int d\Pi_3  \frac{1}{4}\sum_{\substack{\text{pols} \\ \text{cols}}} &  \bigl\lvert\cM( p_1+p_2 \rightarrow k_1+k_2+k_3) \bigr\rvert^2 (2\pi)^4\delta^4(q -\sum_i k_i) \\[-1em]
&\qquad\times \delta_J\bigl(e_2^{(\alpha)} - e_2^{(\alpha)}\{k_1,k_2,k_3\}\bigr)\,, \nn
\end{align}
where $q = p_1 + p_2$ is the total incoming momentum from the $e^+$ and $e^-$, with $q = (Q,0,0,0)$ in the CM frame, $k_{1,2}$ are the momentum of the outgoing $b$ and $\bar b$ quarks, respectively, and $k_3$ the momentum of the outgoing gluon. As usual we average over incoming spins and sum over final spins and colors.
In \eq{fixedordercs}, $\delta_J$ is a delta function imposing the restrictions due to the measurement on the groomed jet in the final state, whose complete form we will work out below.
Without loss of generality, we can assume the unmeasured jet (in this case $k_2$) to be along the $-z$ direction. We then need the other two final state particles to have $k^z_i >0$ so that they are in the right hemisphere. We need both of these particles to have energy greater than $z_\text{cut} Q/2$. The angle between $k_1$ and $k_3$ should be less than $R$. Altogether these constraints determine $\delta_J$ to take the form:
\begin{align}
\delta_J\bigl(e_2^{(\alpha)} - e_2^{(\alpha)}\{k_1,k_2,k_3\}\bigr) &\equiv \delta\Biggl(e_2^{(\alpha)} - \frac{E_1 E_3}{E_J^2}  \biggl(\frac{2k_1\cdot k_3}{E_1 E_3}\biggr)^{\alpha/2}\Biggr) \theta\Bigl(\min(k_1^0,k_3^0) - z_\text{cut} \frac{Q}{2}\Bigr)   \\
&\quad\times \theta(k_1^z)\theta(k_3^z)\theta(\cos\vartheta_{13} > \cos R)\,. \nn
\end{align} 

Meanwhile, the phase space integration measure in \eq{fixedordercs} is given by:
\be
d\Pi_3 = d^4 k_1\, \delta^+(k_1^2 - m_b^2)\, d^4 k_2\, \delta^+(k_2^2 - m_b^2)\, d^4 k_3\, \delta^+(k_3^2)\,.
\ee
Finally, the spin-averaged squared amplitude in \eq{fixedordercs} is given by
\begin{align}
\label{eq:Msquared}
 \frac{1}{4}\sum_{\text{pols,cols}}  \bigl\lvert\cM \bigr\rvert^2 &= \frac{512\pi^3\alpha_\text{em}^2 Q_b^2 \alpha_s C_F N_C}{3Q^2}  \biggl\{ \frac{x_1^2 + x_2^2}{(1-x_1)(1-x_2)} \\
& \quad - \frac{2m_b^2}{Q^2} \biggl[ \frac{1}{(1\minus x_1)^2} + \frac{1}{(1\minus x_2)^2} + \frac{2x_3}{(1\minus x_1)(1\minus x_2)}\biggr] - \frac{4m_b^4}{Q^4} \frac{x_3^2}{(1\minus x_1)^2(1\minus x_2)^2}\biggr\} \nn
 \,,
\end{align}
where $Q_b = -1/3$ is the $b$-quark electric charge, and we have defined $x_i = 2E_i / Q$. Momentum conservation imposes $x_1+x_2+x_3 = 2$. This expression \eq{Msquared} is exact in the quark mass $m_b$. 

Performing as many of the phase space integrals in \eq{fixedordercs} analytically as possible, we obtain the cross section in the form:
\bea
\label{eq:fint}
 \frac{1}{\sigma_0}\frac{ d\sigma}{d e_2^{(\alpha)}} &=& \frac{\alpha_s}{2\pi} C_F \int dx_1 dx_3 \Bigg \{ \frac{x_1^2+ x_2^2}{(1-x_1)(1-x_2)}- \frac{4m^4}{Q^4}\frac{x_3^2}{(1-x_1)^2 (1-x_2)^2}\\
&&-  \frac{2m^2}{Q^2} \biggl[ \frac{1}{(1\minus x_1)^2} + \frac{1}{(1\minus x_2)^2} + \frac{2x_3}{(1\minus x_1)(1\minus x_2)}\biggr] \Bigg \}\nn\\
&&\times\delta \left( e_2^{(\alpha)} - \frac{x_1x_3}{(x_1+x_3)^2} \left( \frac{4(1-x_2)}{x_1x_3}\right)^{\alpha/2}\right)\theta_J(x_1,x_3) \Biggr\rvert_{x_2 = 2-x_1-x_3}  \,, \nn
\eea
where $\sigma_0 = 4\pi\alpha_\text{em}^2 Q_b^2 N_C / (3Q^2)$.
The first term in \eq{fint} is exactly the same as that of the massless jet. The second and third terms may appear as power corrections in $m^2/Q^2$ but they also contribute to the singular cross section. As indicated, $x_2$ in the integrand is function of $x_{1,3}$, and $\theta_J$ is the set of further constraints imposed by the jet algorithm and jet grooming, which we will now write out more explicitly.

The soft drop condition is implemented as 
\bea
\label{eq:softdropconstraint}
\frac{\text{min}[x_1,x_3]}{x_1+x_3} > z_\text{cut}.
\eea
Meanwhile, the (geometric) angle between $k_1$ and $k_3$ is given by
\bea
\cos \vartheta_{13} = \frac{1-x_1-x_3 +x_1x_3/2}{x_3 \sqrt{x_1^2/4-m_b^2/Q^2}} \,.
\eea
The lower limit is set by $\vartheta_{13} =0$, which is the collinear limit and we need the mass to be nonzero to regulate this collinear divergence. Since $\cos \vartheta_{13} < 1$ so,
\begin{equation}
\label{eq:costheta}
 \frac{1-x_1-x_3 +\frac{x_1x_3}{2}}{x_3 \sqrt{\frac{x_1^2}{4}-\Delta}} \leq 1,
\end{equation}
where we have defined $\Delta=m_b^2 /Q^2$.
This equation tells us that the $\Delta$ factor under the square root prevents the singularity $x_1+x_3 =1$ or $x_2 =1$. We can rewrite \eq{costheta} as, 
\bea
\label{eq:angle}
x_3 \geq \frac{1-x_1}{1-\frac{x_1}{2}+ \sqrt{\frac{x_1^2}{4}-\Delta}}.
\eea
\begin{figure}
\vskip-.1cm
\centerline{\scalebox{.5}{\includegraphics{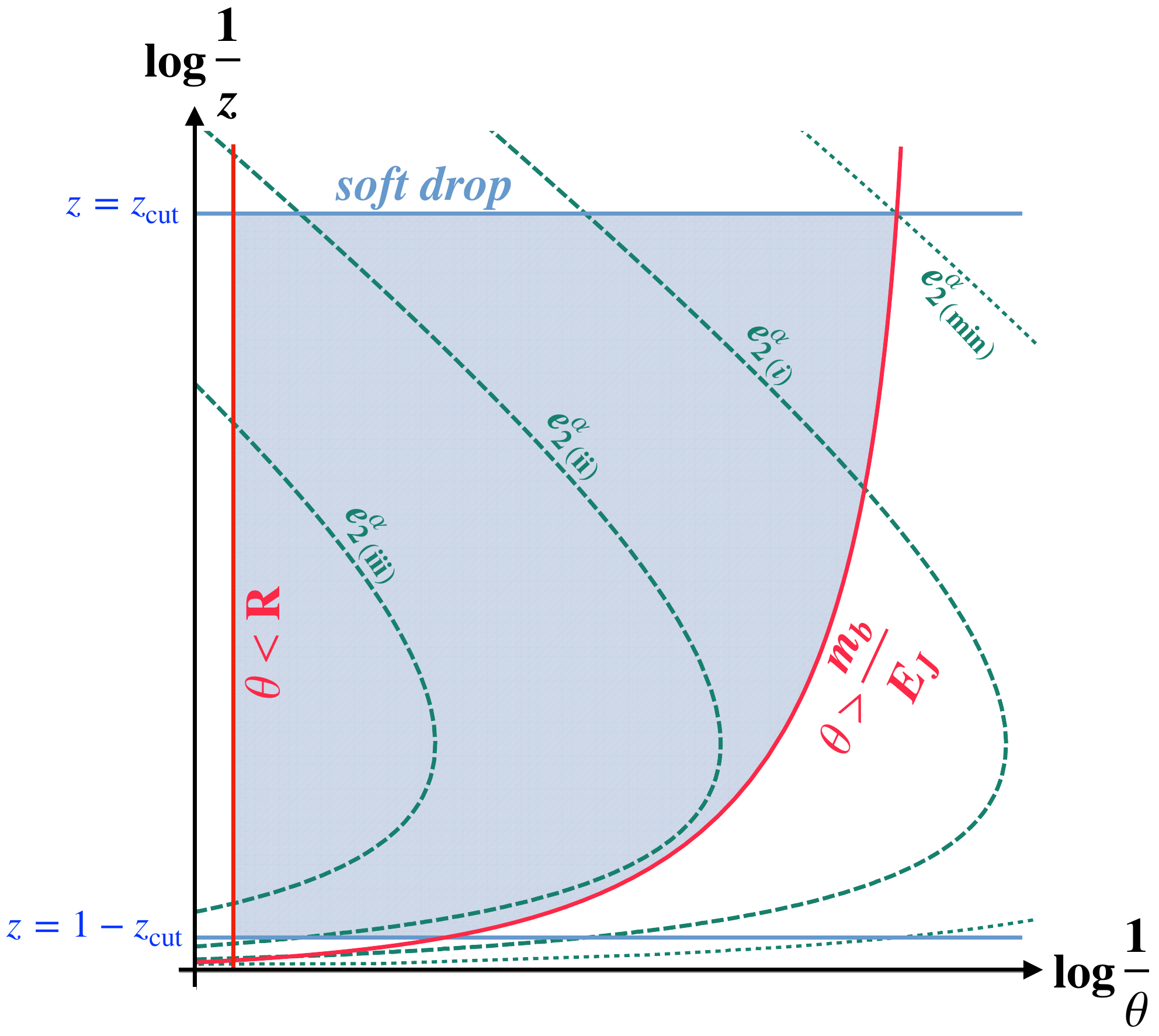}}}\vskip-0.3cm
\caption[1]{The phase space for integration in the $\ln(1/\theta),\ln(1/z)$ plane at $\cO(\as)$ in full QCD. The boundaries are determined by the jet radius ($R$), the soft drop condition ($z_\text{cut}$), and the minimum angle at which a quark of nonzero mass $m_b$ and energy $E_J$ can emit radiation. The allowed phase space region is shaded. The dotted line is the smallest allowed value $e_{2,\text{min}}^{(\alpha)}$ due to this minimum angle. The dashed lines are lines of constant $e_2^{(\alpha)}$, for (i) a small value where $e_2^{(\alpha)}$ is sensitive to both quark mass and grooming, and the EFT calculation is applicable; (ii) an intermediate value where it is sensitive to  grooming but is just above the quark mass constraint, causing the first kink in the fixed-order cross section; and finally (iii) a large value where it moves past the grooming boundaries and is affected only by the jet radius, causing the second kink in the fixed-order cross section. The region where resummation is important is near the corner of the quark mass and grooming boundaries in region (i), approaching the collinear and soft divergences. In this region the phase space goes to the parametric limit shown in \fig{ztheta} and factors into the EFT regions shown there.}
\label{fig:phase} 
\end{figure}
The other limit is set by $\vartheta = R $ which is not a region of any singularity, hence we can drop $\Delta$ entirely in this case:
\be
\label{eq:Radius}
 \frac{2(1-x_1-x_3 )+x_1x_3}{x_3 x_1} \geq\cos R \quad\Rightarrow\quad x_3 \leq \frac{1 - x_1} {1-x_1 \sin^2 \frac{R}{2}} \,.
\ee
The only other restriction which regulates soft singularities is the soft drop condition which gives us 
\begin{equation}
\label{eq:sd}
 x_3 \geq \frac{x_1 z_\text{cut}}{1- z_\text{cut}},~~~~~~x_3 \leq \frac{x_1(1- z_\text{cut})}{ z_\text{cut}}.
\end{equation}
These four conditions in \eqss{angle}{Radius}{sd}, define the theta function $\theta_J$ in \eq{fint} and define the boundaries of our phase space, which are illustrated in \fig{phase}. For this illustration we have found it more convenient to visualize in the variables:
\be
\label{eq:ztheta}
z = \frac{x_3}{x_1+x_3} \Rightarrow 1-z = \frac{x_1}{x_1+x_3}\,,\qquad \theta^2 = \frac{4(x_1+x_3-1)}{x_1 x_3}\,,
\ee
and to plot in $\ln(1/\theta)$ and $\ln(1/z)$ as in \figs{ztheta}{ztheta2}, which themselves represent the parametric behavior of \fig{phase} in the soft and collinear limits.

\begin{figure}
\centerline{\scalebox{.6}{\includegraphics{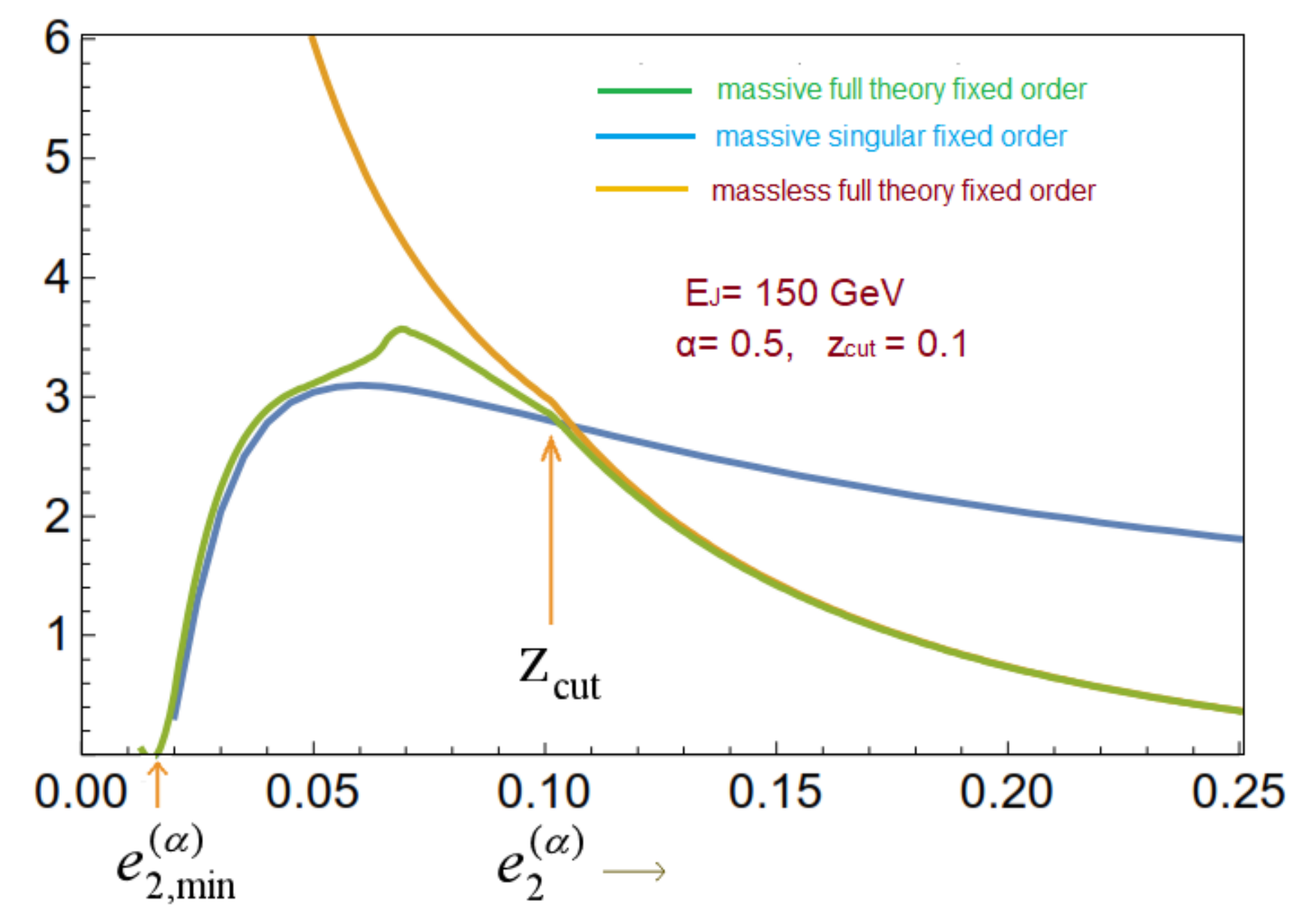}}}
\vskip-0.5cm
\caption[1]{Fixed-order cross section at one loop for the $e_2^{(\alpha)}$ distribution. The massive full theory fixed-order cross section (green) is computed numerically by integration over the phase space in \fig{phase} and has kinks in the shape which corresponding to  physical scales where quark mass or grooming constraints are crossed. The massive singular distribution is in blue (which is regulated by the quark mass itself) while the massless full theory fixed-order prediction is shown in orange.}
\label{fig:Fixed} 
\end{figure}

We compute the phase space integral  numerically to give the full theory fixed-order cross section.
We do this by integrating over the whole allowed area of phase space illustrated in \fig{phase} above each line of constant $e_2^{(\alpha)}$, and differentiating the result with respect to $e_2^{(\alpha)}$ to obtain the differential cross section \eq{fint}.
Looking at the result in \fig{Fixed}, we see that there are  a number of turning points/kinks in the fixed-order cross section and it is instructive to try and understand how these points arise and what scales they correspond to.
Given the physical constraints, we can get an intuitive understanding about why these kinks exist.
It is clear that due to the energy cutoff imposed by soft drop and the angular cutoff imposed by the mass of the heavy quark, we have a minimum value of $e_2^{(\alpha)}$ below which the cross section is zero. Since the heavy quark carries $\mathcal{O}(1)$ of the jet energy, we should expect 
\bea
e_{2, \text{min}}^{(\alpha)} = z_\text{cut}(1-z_\text{cut})^{1-\alpha} \left(\frac{m}{E_J}\right)^{\alpha}.
\eea 
In the singular limit, we keep the leading term up to power correction in $z_\text{cut}$, which gives us the scale 
$e_{2, \text{min}}^{(\alpha)} \approx z_\text{cut} ({m}/{E_J})^{\alpha}$.

In \fig{Fixed} we see there are two kinks at larger $e_2^{(\alpha)}$ in the fixed-order cross section. From \fig{phase} we can see that the first of these kinks happens when the $e_2^{(\alpha)}$ distribution becomes insensitive to the mass of the heavy quark. The EFT power counting  tells us that at all orders in perturbation theory, this scale should be the parametrically of the order $(m/E_J)^{\alpha}$. However, at the first trivial order in perturbation theory, where we have a single gluon emission off the heavy quark, we see a kink in the fixed-order cross section at a much lower value of $e_2^{(\alpha)}$. To determine this value, it is convenient to work in the variables $z,\theta$ defined in \eq{ztheta} and refer to \fig{phase}. In these variables, the lines of constant $e_2^{(\alpha)}$ are determined by
\be
e_2^{(\alpha)} = z(1-z)\theta^\alpha\,.
\ee
We want to determine the value of $e_2^{(\alpha)}$ for which this line lies entirely above the constraint (red line in \fig{phase}),
\be
\theta>\theta_\text{min} = \frac{2\sqrt{\Delta}}{1-z}\,.
\ee
This happens if and only if
\be
\label{eq:e2maxcondition}
e_2^{(\alpha)} > (4\Delta)^{\alpha/2} z(1-z)^{1-\alpha}\,,
\ee
for all $z$ within the allowed region $z_\text{cut}<z<1-z_\text{cut}$.  The maximum value that $f(z) = z(1-z)^{1-\alpha}$ can take in this region can be determined by solving the simple maximization condition,
\be
\frac{df}{dz} = 0 \Rightarrow z = \frac{1-\alpha}{2-\alpha}\,,
\ee
which determines the largest value that the right-hand side of \eq{e2maxcondition} can be, and thus determines the minimum value of $e_2^{(\alpha)}$ for which the cross sections just escapes the mass constraint line:
\bea
 e_2^{(\alpha)} \geq (4\Delta)^{\alpha/2}  \frac{(1-\alpha)^{1-\alpha}}{(2-\alpha)^{2-\alpha}}.
\eea
For $\alpha = 0.5$ this gives us a value of $e_2^{(\alpha)} \gtrsim 0.07$, which  is indeed where we see the kink in the one-loop fixed-order cross section in \fig{Fixed}. This occurs between regions (i) and (ii) in \fig{phase}. As we go to higher orders, the phase space increases and hence we expect this to get closer to the EFT prediction of $(4\Delta)^{\alpha/2}$.

Finally we consider a configuration where the angle $\theta_{ij} \sim 1$ and the accompanying radiation just passes soft drop ($z \sim z_\text{cut}$), $e_2 ^{(\alpha)} \sim z_\text{cut}$. Beyond this value of $e_2^{(\alpha)}$, we expect that the soft drop condition will no longer be relevant and the jet will behave as a massless ungroomed jet. This gives the second kink in the cross section at $e_2^{(\alpha)} \sim z_\text{cut}$, which occurs between regions (ii) and (iii) in \fig{phase}.

We wish to note again the curious fact that the mass and grooming kinks or transitions in the fixed-order cross section occur in the \emph{opposite} order of the EFT transitions in \fig{ztheta}---for the EFTs, the grooming constraint turns off while the mass is still relevant. This just means that the procedure to predict the singular logs encounters these constraints differently than the computation of the exact one-loop fixed-order power corrections. As noted above, the phase space for the fixed-order calculation grows at higher orders and will approach the behavior of the resummed calculation more closely. For the fixed-order calculation, we have to account for the exact behavior of the phase space boundaries and $e_2^{(\alpha)}$ contours for $z\to 1$ away from the soft limit in which \fig{ztheta} is drawn, leading to the behavior shown in \fig{phase}. That is to say, the kinks in the cross section in \fig{Fixed} are from the fixed-order power corrections, not the singular part of the cross section. As we are about to see, the EFT and full QCD computations of the singular logs themselves agree perfectly in the limit of small $e_2^{(\alpha)}$.


\subsection{Fixed-order collinear limit}
\label{ssec:collinear}

We can compute the cross section at one loop analytically in the collinear (low $e_2^{(\alpha)}$) limit. By comparing the singular terms from the expansion of the resummed cross section and the fixed-order collinear limit, we can fix the normalization for our resummed result in order to do the matching to the full theory one-loop fixed-order result.

Computing the cross section \eq{fint} in this limit is most transparent in terms of the variables $z,\theta$ in \eq{ztheta}.
Solving these relations for $x_{1,3}$, we obtain
\be
x_1 = \frac{2}{\theta^2 z} \Bigl( 1 - \sqrt{ 1 - \theta^2 z(1-z)}\Bigr)\,,\qquad x_3 = \frac{2}{\theta^2 (1-z)} \Bigl( 1 - \sqrt{ 1 - \theta^2 z(1-z)}\Bigr)\,.
\ee
In the collinear limit, we only need these relations in the small $\theta$ limit, 
\be
x_1 \approx (1-z) \Bigl[ 1 + \frac{\theta^2 z(1-z)}{4}\Bigr] \,,\qquad x_3 \approx z \Bigl[ 1 + \frac{\theta^2 z(1-z)}{4}\Bigr]\,,
\ee
which gives the Jacobian in the collinear limit,
\be
dx_1\,dx_3 = \frac{1}{2} z(1-z) dz\,\theta \,d\theta\,.
\ee
Then, the cross section \eq{fint} in this limit becomes
\begin{align}
\frac{1}{\sigma_0}\frac{ d\sigma}{d e_2^{(\alpha)}}  = \frac{\as C_F}{2\pi} &\int_{z_\text{cut}}^{1} dz\int_{2m/Q}^{1}d\theta \, \delta\Bigl( e_2^{(\alpha)} - z(1-z)\theta^\alpha\Bigr)  \biggl( \frac{4}{z\theta} - \frac{16m^2}{Q^2} \frac{1}{z\theta^3}\biggr) + \cdots\,,
\end{align} 
up to terms that will be suppressed by additional powers of $m^2/Q^2$, $z_\text{cut}$, or $e_2^{(\alpha)}$. We perform the $\theta$ integral using the delta function, which in order to have a solution, places a tighter upper limit on the $z$ integral, and also imposes the bound $e_2^{(\alpha)} > e_{2,\text{min}}^{(\alpha)}$, as is clear from \figs{ztheta}{phase}:
\begin{align}
\frac{1}{\sigma_0}\frac{ d\sigma}{d e_2^{(\alpha)}}  = \frac{\as C_F}{2\pi} \frac{4}{\alpha e_2^{(\alpha)}} \theta\bigl(e_2^{(\alpha)} - e_{2,\text{min}}^{(\alpha)}\bigr) \int_{z_\text{cut}}^{e_2^{(\alpha)}/\theta_\text{min}^\alpha} \frac{dz}{z}\biggl[1 - \frac{4m^2}{Q^2} \biggl( \frac{z}{e_2^{(\alpha)}}\biggr)^{2/\alpha}\biggr]\,,
\end{align}
up to terms suppressed by additional powers of $z_\text{cut}$ or $e_2^{(\alpha)}$. The final result of this integral is then
\be
\label{eq:collinearFO}
\frac{1}{\sigma_0}\frac{ d\sigma}{d e_2^{(\alpha)}}  = \frac{\as C_F}{2\pi}\theta\bigl(e_2^{(\alpha)} - e_{2,\text{min}}^{(\alpha)}\bigr) \Biggl\{\frac{4}{\alpha} \frac{1}{e_2^{(\alpha)} }\ln \frac{e_2^{(\alpha)}}{e_{2,\text{min}}^{(\alpha)}} - \frac{2}{e_2^{(\alpha)}} \Biggl[ 1 - \biggl( \frac{e_{2,\text{min}}^{(\alpha)}}{e_2^{(\alpha)}}\biggr)^{2/\alpha}\Biggr]\Biggr\}\,,
\ee
where, again,
\bea
e_{2,\text{min}}^{(\alpha)} = z_\text{cut} \left(\frac{2m}{Q}\right)^{\alpha}  \,.
\eea
This cross section has large logs in $e_2^{\alpha}$, for low $e_2^{(\alpha)}$, which are cut off at $e_{2,\text{min}}^{(\alpha)}$. We see explicitly double and single logs in $e_2^{(\alpha)}$ (where $1/e_2^{(\alpha)}$ counts as a log), where grooming and quark mass have replaced one $\ln e_2^{(\alpha)}$ that would be present for ungroomed massless jets with a smaller log of $\ln (e_2^{(\alpha)}/e_{2,\text{min}}^{(\alpha)})$.
This is typical of grooming where one of the logs in the double logarithm of the measurement gets replaced by a smaller log associated with grooming and/or any relevant mass parameter, see e.g.~\cite{Larkoski:2014wba} or \cite{Makris:2017arq} where the  a log of the energy correlator $C_1^{(\alpha)}$ or transverse momentum measurement $q_{\perp}$ gets replaced by $\ln z_{\text{cut}}$.We also have double and single logarithms in $\ln(E_J/m)$ from the virtual diagrams which get resummed at NLL.

Our EFT reproduces this cross section in the collinear limit at one loop order, which is a powerful cross-check on the validity of our EFT calculation. Note that for $e_2^{(\alpha)}\gg e_{2,\text{min}}^{(\alpha)}$, the last term is power suppressed, and the $e_2^{(\alpha)}$ dependence of the other two terms is reproduced by the combination of the leading-order collinear-soft and ultracollinear functions $S_C$  and $B_+$. At the low endpoint $e_2^{(\alpha)}\to e_{2,\text{min}}^{(\alpha)}$, as illustrated in \fig{ztheta}, the two scales merge, the last term in \eq{collinearFO} becomes leading order, and the result is reproduced instead by  $B_+^{SD}$ in \eq{bjetSD} at nonzero values of $e_2^{(\alpha)}$.
We plot out the singular cross section in \fig{Fixed} and see that in the low $e_2^{(\alpha)}$ regime, it does a good job of describing the fixed-order $e_2^{(\alpha)}$ distribution.


\section{Results at NLL}
\label{sec:resum}

To compute the cross section at next-to-leading-logarithmic (NLL) accuracy,  i.e. up to and including terms of order $\as^n \ln^n s$ in Laplace space\footnote{To be precise, the double logs in this hierarchy are $\ln s \, \ln (se_{2\text{min}}^{(\alpha)})$, see \eq{collinearFO}. We also sum double and single logs of $m/E_J$ at NLL, which affect the normalization but not shape of the $e_2^{(\alpha)}$ distribution, see below.}, see e.g. \cite{Almeida:2014uva}, we need the one- and two-loop cusp anomalous dimensions as well as the one-loop non-cusp anomalous dimensions. The required one-loop results have been given in \sec{oneloopEFT}. The two-loop cusp anomalous dimension can be obtained from the literature \cite{Korchemsky:1987wg}. This gives us all the ingredients for NLL resummation. We will concentrate on the regime $ z_\text{cut} \sim (m/E_J)^{\alpha} \gg e_2^{(\alpha)}$. This depends on the value of $\alpha $ and $\omega$, but is typically true for $\alpha <1$.
Thus we will only need to make use of the ``region I'' and ``region 0''  EFT described in \ssec{EFTregions}.

As we move towards larger values of $e_2^{(\alpha)}$, we reach a transition region in which we must smoothly match to the full theory fixed-order cross section which includes power corrections in $e_2^{(\alpha)}/(4\Delta)^{\alpha/2}$ and  $e_2^{(\alpha)}/z_\text{cut}$ where $\Delta\equiv m_q^2/E_J^2$. This is done using a profile function that smoothly turns off resummation in the transition region. At the same time, for the regime of very small $e_2^{(\alpha)} \sim e^{(\alpha)}_{2,\text{min}}$, we need to smoothly transition to the  ``region 0'' EFT described in \ssec{EFTregions}, matching onto power corrections in $e_2^{(\alpha)}/e_{2,\text{min}}^{(\alpha)}$, again using a profile function for the renormalization scale.


\subsection{Resummation at NLL}

Consider a function with an anomalous dimension $\gamma$ which can be written as, 
\bea
\gamma = \Gamma_F(\alpha_s) \ln \left(\frac{\mu^2}{\mu_F^2}\right)+ \gamma_F(\alpha_s), 
\eea
where $\alpha_s$ is the strong coupling constant, $\Gamma_{F}$ is the cusp anomalous dimension and $\gamma_{F}$ is the non-cusp anomalous dimension. We expand these anomalous dimensions as a series in $\alpha_s$.
\bea
\Gamma_F &=& C_i \Gamma_\text{cusp}, \ \ \ \ \  \Gamma_\text{cusp} = \sum_{n=0}^{\infty} \Gamma_n \left(\frac{\alpha_s}{4\pi}\right)^{n+1} \,,\qquad\gamma_F= \sum_{n=0}^{\infty} \gamma_n \left(\frac{\alpha_s}{4\pi}\right)^{n+1} 
\eea
where $C_i$ is the appropriate color factor. The one loop ($\Gamma_0$) and two loop ($\Gamma_1$) cusp anomalous dimension are given by \cite{Korchemsky:1987wg}, 
\bea 
 \Gamma_0 &=& 4 \,,\qquad \Gamma_1 = 4C_A \left(\frac{67}{9}-\frac{\pi^2}{3}\right) -\frac{80}{9}T_Rn_f\,.
\eea

\begin{figure}
\centerline{\scalebox{.4}{\includegraphics{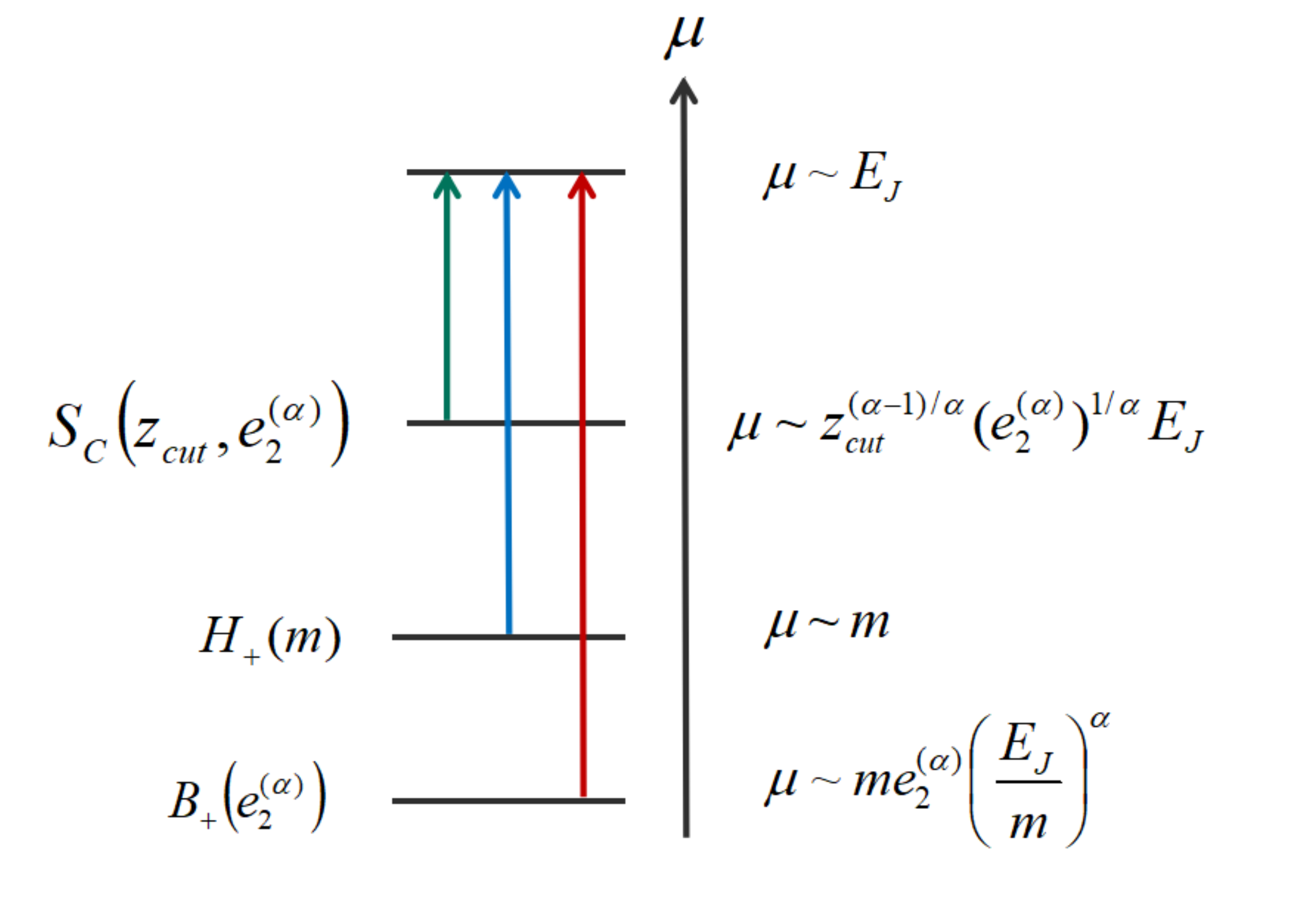}}}
\vskip-0.5cm
\caption[1]{Resummation path evolving ultracollinear, HQET matching, and collinear-soft functions from their natural scales to the common scale $\mu\sim E_J$, where the hard function is evaluated.}
\label{fig:resum} 
\end{figure}

The resummed result for a function $F$ at the scale $\mu$ to NLL accuracy can be written down as
\bea
F(\mu) =  e^{K_F(\mu,\mu_0)} F(\mu_0) \left(\frac{\mu_0^2}{\mu_F^2}\right)^{\omega_F(\mu,\mu_0)} 
\eea
where we have evolved the function from $\mu_0$ to $\mu$ with 
\be
K_F(\mu,\mu_0) =2\int_{\mu_0}^\mu \frac{d\mu'}{\mu'} \Gamma_F[\as(\mu')]\ln\frac{\mu'}{\mu_0}\,,\quad \omega_F(\mu,\mu_0) = \int_{\mu_0}^\mu \frac{d\mu'}{\mu'} \Gamma_F[\as(\mu')]\,,
\ee
which have expansions up to NLL accuracy,
\bea
K_F &=& \frac{C_i \Gamma_0}{2 \beta_0^2} \left(\frac{4 \pi}{\alpha_s(\mu_0)} \left(\ln r +\frac{1}{r} -1\right)+\left(\frac{\Gamma_1}{\Gamma_0}-\frac{\beta_1}{\beta_0}\right)(r-1-\ln r)-\frac{\beta_1}{2\beta_0} \ln^2 r\right) -\frac{\gamma_0}{2\beta_0}\ln r \nn\\
\omega_F &=& -\frac{C_i\Gamma_0}{2\beta_0} \left(\ln r + \frac{\alpha_s(\mu_0)}{4\pi} \left(\frac{\Gamma_1}{\Gamma_0}-\frac{\beta_1}{\beta_0}\right)(r-1)\right)
\label{eq:KFomegaF}
\eea
and $ r = \alpha_s(\mu)/\alpha_s(\mu_0)$.

\begin{enumerate}

\item{Region I}\\
We have to run three functions ($S_C$, $B_{+}$, $H_+$) defined in \sec{oneloopEFT}, from their natural scales to the hard scale $E_J$. The path for resummation is shown in \fig{resum}. The collinear-soft function $S_C$ is the same as for the massless case and so its resummation is the same as in Ref.~\cite{Frye:2016aiz}.
We do the scale setting in $e_2^{(\alpha)}$ space. For the three functions that we consider, we write down their natural scales and anomalous dimensions. 
\begin{itemize}
\item {Collinear-Soft}
\bea
\mu_{CS} = E_J z_\text{cut} \left(s z_\text{cut}\right)^{-1/\alpha} \,, \qquad C_i = -\frac{C_F \alpha}{\alpha-1} \,,\qquad 
\gamma_0 = 0
\eea
\item {HQET matching function}
\bea
\mu_{H} = m \,, \qquad C_i= C_F \,, \qquad \gamma_0 = 2C_F
\eea
\item{HQET jet function}
\bea
\mu_{B_+} = \frac{E_J}{s (4\Delta)^{(1-\alpha)/2}} \,, \qquad C_i = \frac{C_F}{\alpha-1}\,, \qquad \gamma_0 = 4C_F \,.
\eea
\end{itemize}
Then the resummed HQET jet function can be written as 
\bea
\widetilde B_+(s, \mu) = e^{K_{B_+}(\mu, \mu_0^{B_+})} \widetilde B_+( L_C \rightarrow \partial_{\omega_{B_+}})\left[\frac{(\mu_0^{B_+})^2}{\mu^2_{B_+}}\right]^{\omega_{B_+}(\mu, \mu_0^{B_+})}\,.
\eea
The argument of $\widetilde B_+$ on the right-hand side indicates that its natural argument, $L_C$ defined in \eq{LCLV}, is to be replaced by the derivative operator $\partial_{\omega_{B_+}}$, which generates the same logarithms when acting on the evolution kernel on the right (see \cite{Becher:2006mr,Becher:2006nr,Almeida:2014uva}).
A similar formula will hold for the collinear-soft function.
Then the final result for the cross section in Laplace space can be written as, 
\bea
 \sigma(s) &=& \left(\frac{(\mu_0^{H_+})^2 }{m^2}\right)^{\!\omega_{H_+}\!(\mu, \mu_0^{H_+}\!)}\! \exp\left[K_{H_+}(\mu,\mu_0^{H_+}\!)+K_{CS}(\mu, \mu_0^{CS})+K_{B_+}(\mu, \mu_0^{B_+}\!)\right] \\
&&\times B_+( L_C \rightarrow \partial_{\omega_{B_+}}\!)\left[\frac{(\mu_0^{B_+}\!)^2}{\mu^2_{B_+}}\right]^{\omega_{B_+}(\mu, \mu_0^{B_+}\!)}S_C( L_{S_C} \rightarrow \partial_{\omega_{CS}})\left[\frac{(\mu_0^{CS})^2}{\mu^2_{CS}}\right]^{\omega_{CS}(\mu, \mu_0^{B_+})} \,. \nn
\eea

Since the resummation of $B_+$ and $S_C$ involves the Laplace variable $s$, and we need to go back to $e_2^{(\alpha)}$ space. To that end we can use 
\bea
\mathcal{L}^{-1}(s^q)= \frac{(e_2^{(\alpha)})^{-q-1}}{\Gamma[-q]} \,.
\eea

We can now write the resummed cross section in $e_2^{(\alpha)}$ space 
\begin{align}
\label{eq:NLLresult}
&\frac{1}{\sigma_0(z_\text{cut},E_J)}\frac{d \sigma}{d e_2^{(\alpha)}} =\frac{1}{e_2^{(\alpha)}} \left(\frac{(\mu_0^{H_+})^2 }{m^2}\right) ^{\omega_{H_+}(\mu, \mu_0^{H_+})}e^{K_{H_+}(\mu,\mu_0^{H_+})+K_{CS}(\mu, \mu_0^{CS})+K_{B_+}(\mu, \mu_0^{B_+})} \nn \\
&\times S_C( L_{S_c} \rightarrow \partial_{\omega_{CS}})\left(\frac{(\mu_0^{CS})^2(e_2^{(\alpha)}e^{-\gamma_E})^{-2/\alpha}}{E_J^2 (z_\text{cut})^{2(\alpha-1)/\alpha}}\right) ^{\omega_{CS}(\mu, \mu_0^{CS})} \\
&\times B_+( L_C \rightarrow \partial_{\omega_{B_+}})\left(\frac{(\mu_0^{B_+})^2 (4\Delta)^{\alpha-1}}{(e_2^{(\alpha)}e^{-\gamma_E})^2 E_J^2}\right) ^{\omega_{B_+}(\mu, \mu_0^{B_+})}\frac{1}{\Gamma\left(-\frac{2}{\alpha}\omega_{CS}(\mu, \mu_0^{CS})-2\omega_{B_+}(\mu, \mu_0^{B_+})\right)}  \,. \nn
\end{align}
Since we are working at NLL accuracy, we keep all the terms of $\cO(\alpha_s L \sim 1)$.
We wish to minimize the logs so we make the choice 
\begin{align}
\mu_0^{B_+}& = e_2^{(\alpha)} \omega \Delta^{(1-\alpha)/2}= m \frac{e_2^{(\alpha)}}{(4\Delta)^{\alpha/2}}\nn\\
\mu_0^{H_+} &= m \nn\\
 \mu_0^{CS} &= E_J z_\text{cut} \left(\frac{e_2^{(\alpha)}}{z_\text{cut}} \right)^{1/\alpha}
\end{align}
with the scale $\mu$ is set to $E_J$.\footnote{Technically \eq{NLLresult} should be independent of $\mu$, and is when considered to all orders, but because of the way the expansion of $K_F$ in \eq{KFomegaF} is carried out, retains a small, subleading $\mu$ dependence. See \cite{Bell:2018gce} for explanation and an alternative expansion that removes this residual $\mu$ dependence exactly at every order.} We will vary around these default choices later to estimate theoretical uncertainty.

\item{Region 0} \\
We have to run two functions ($B^{SD}_{+}$, $H_+$) defined in \sec{oneloopEFT}, from their natural scales to the hard scale $E_J$. The path for resummation is shown in \fig{resum}. The resummation for the $H_+$ function is identical to that of Region I. For the soft drop constrained HQET jet function, the natural scale is simply $mz_\text{cut}$ which is independent of the measurement scale $e_2^{(\alpha)}$ (or $s$ in Laplace space) so that 
\bea
 \mu_{B_+^{SD}} =mz_\text{cut} \,, \qquad C_i = -C_F\,, \qquad \gamma_0 = 4C_F \,.
\eea

Following the same process as for Region I, we write the resummed cross section in $e_2^{(\alpha)}$ space as 
\begin{align}
\label{eq:NLLresult3}
\frac{1}{\sigma_0(z_\text{cut},E_J)}\frac{d \sigma}{d e_2^{(\alpha)}} &=e^{K_{H_+}(\mu,\mu_0^{H_+})+K_{B_+^{SD}}(\mu, \mu_0^{B_+^{SD}})}\nn\\
&\times \left(\frac{(\mu_0^{H_+})^2 }{m^2}\right)^{\omega_{H_+}(\mu, \mu_0^{H_+})}\left(\frac{(\mu_0^{B_+})^2 }{m^2z^2_\text{cut}}\right) ^{\omega_{B_+^{SD}}(\mu, \mu_0^{B_+^{SD}})} B_+^{SD}(e_2^{(\alpha)})\,,
\end{align}
where $B_+^{SD}$ is evaluated in a fixed-order expansion, given to one loop in \eq{bjetSD}.

As before $\mu \sim E_J$ and to minimize the logarithms, we set 
\bea
\mu_0^{H_+} = m\,,  \qquad  \mu_0^{B_+^{SD}} = mz_\text{cut} \,.
\eea
It is clear from this form that the resummation in this region is only acting as an overall normalization factor. The shape of the $e_2^{(\alpha)}$ distribution is given entirely by the fixed-order result. However, it is still necessary to keep the normalization factor so as to smoothly match to the Region I result.  Moreover,  the resummation exponent in Region 0 can be obtained from that in Region I simply by setting $\mu_0^{B_+} = \mu_0^{CS} = mz_\text{cut}$.
\end{enumerate}


\subsection{Partonic $e_2^{(\alpha)}$ spectrum}

To obtain the distribution over the full range of $e_2^{(\alpha)}$, we need to smoothly match Region 0, Region I and the full theory fixed-order cross section in the tail. In order to do that, we need to have profile scales that alter or turn off resummation factors while making a transition from one region to another. 
\begin{itemize}
\item{Region 0 -- Region I}\\
Region 0 is a small region around $e_2^{(\alpha)} \sim e^{(\alpha)}_{2,\text{min}}$. Hence we choose a point $e_2^{(\alpha)} \sim 2 e^{(\alpha)}_{2,\text{min}} $ to transition from region I to Region 0. This is achieved by smoothly taking the scales $\mu_0^{B_+}$ and $\mu_0^{CS}$ to $mz_\text{cut}$ around $e_2^{(\alpha)} \sim 2 e^{(\alpha)}_{2,\text{min}}$. This then automatically matches to the Region 0 cross section except for the power correction term $ -\frac{\alpha_s C_F}{\pi} \theta\bigl(e_2^{(\alpha)} - e_{2,\text{min}}^{(\alpha)}\bigr)\frac{1}{e_2^{(\alpha)}}\left(e^{(\alpha)}_{2,\text{min}}/e_2^{(\alpha)}\right)^{2/\alpha}$. This can be added to the cross section by matching to the fixed-order $B_+^{SD}$ function \eq{bjetSD} at low $e_2^{(\alpha)}$ to ensure a complete matching between the two regions. This is like the fixed-order matching to the full QCD cross section in the tail later.

\begin{figure}
\centerline{\scalebox{.55}{\includegraphics{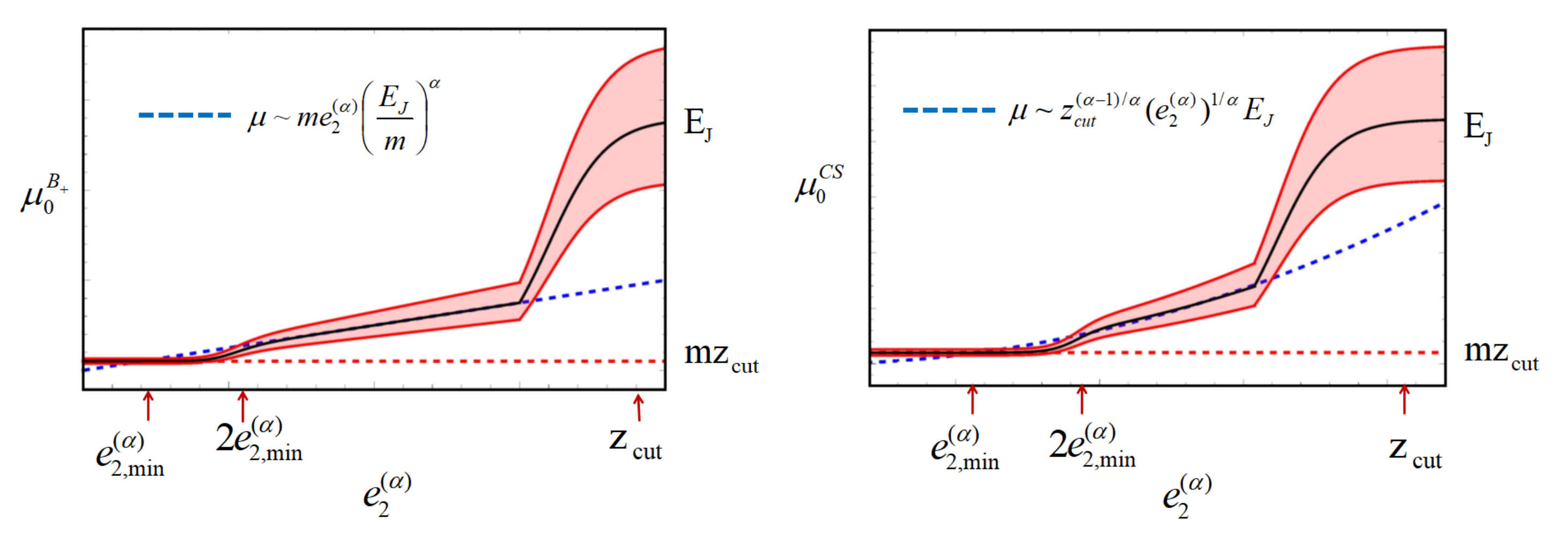}}}
\vskip-0.5cm
\caption[1]{Profiles in the renormalization scales for the bHQET jet (left) and the collinear-soft (right) functions. These two functions combine into a single one ($B_+^{SD}$) for $e_2^{(\alpha)} \sim e_{2,\text{min}}^{(\alpha)}$ which has a natural scale $\mu = mz_{\text{cut}}$. For $e_2^{(\alpha)}\geq z_{\text{cut}}, (m/E_J)^{\alpha}$, the resummation is turned off by setting $\mu = E_J$. The pink band indicates a scale variation by a factor of 2 above and below the central profile. }
\label{fig:profile} 
\end{figure}

\item{Region I -- Tail }\\
For $e_2^{(\alpha)} \geq z_\text{cut} , (m/E_J)^{\alpha}$, we need to turn off the resummation and make a transition to the full theory fixed-order cross section. To do that, we turn off all resummation by smoothly taking all the IR scales $\mu_0^{B_+}, \mu_0^{CS}, \mu_0^{H_+}$ to $E_J$. This then leaves behind the singular part of the fixed-order cross section. The matching is completed by adding in the power correction, i.e., the difference between the singular and the full theory fixed-order cross section.
\end{itemize}
The profile scales for the HQET jet function $B_+$ and the collinear-soft function $S_C$, which enable us to implement this matching between various regions are shown in \fig{profile}.
We implement these profile scales piecewise for the transition between the various regions.
For the region 0--I transition we can employ a suitably modified hyperbolic tangent function. For example for $\mu_0^{B_+}$ we can choose
\begin{figure}
\centerline{\scalebox{.55}{\includegraphics{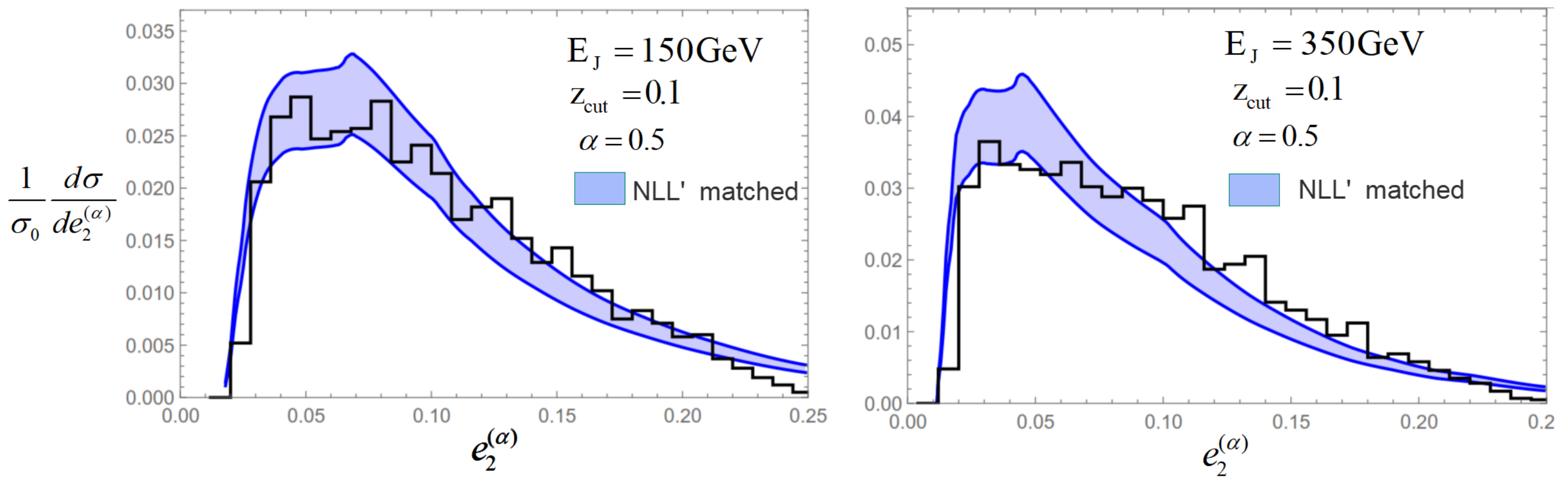}}}
\vskip-0.4cm
\caption[1]{Resummed cross section at NLL$'$ accuracy matched onto one loop fixed-order cross section.}
\label{fig:bplots} 
\end{figure}
\begin{figure}
\centerline{\scalebox{.32}{\includegraphics{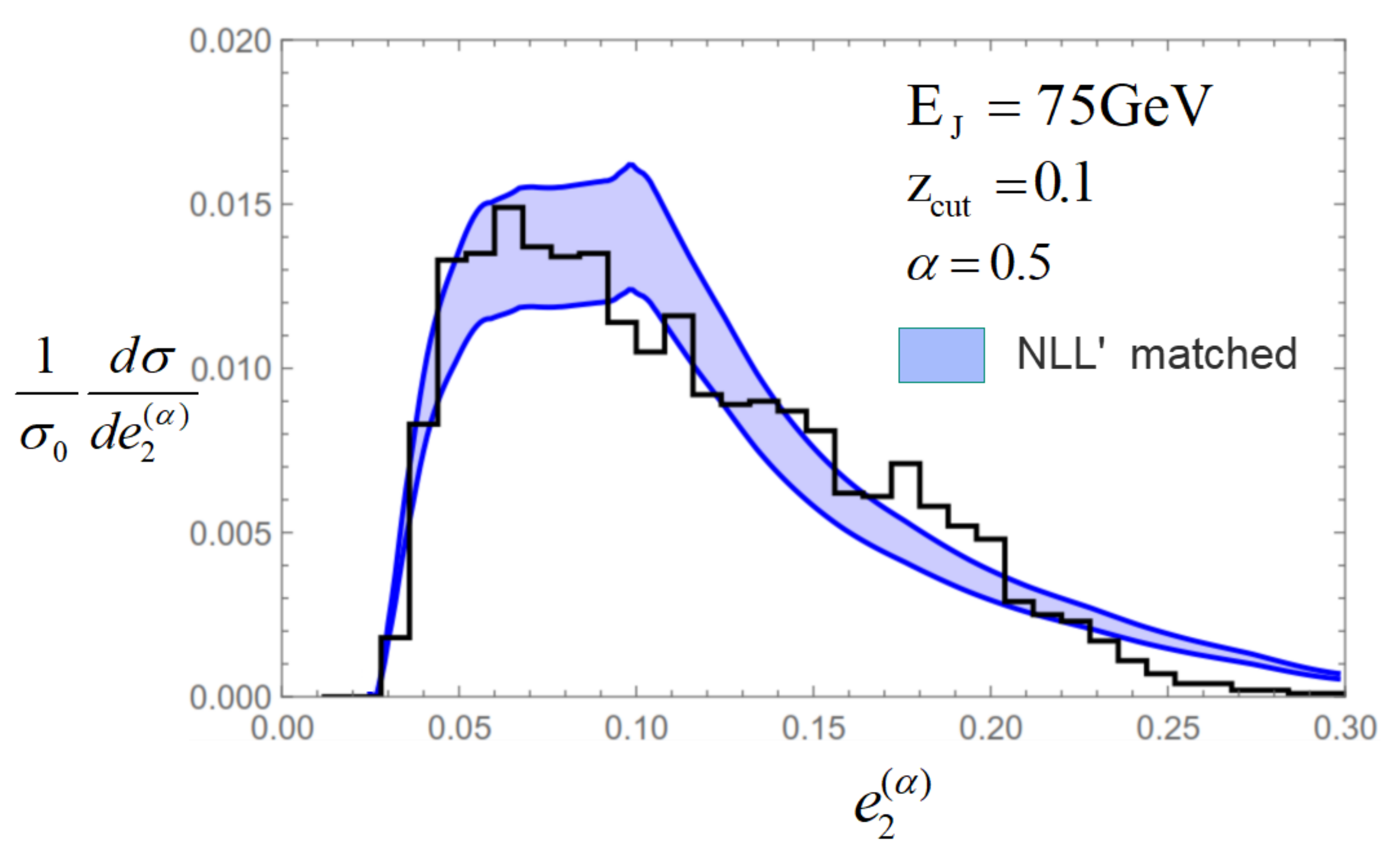}}}
\vskip-0.4cm
\caption[1]{Resummed cross section at NLL$'$ accuracy matched onto one loop fixed-order cross section for jet energy of 75 GeV.}
\label{fig:bplots75} 
\end{figure}
\bea
\mu_0^{B_+}(e_2^{(\alpha)}) =  m z_\text{cut} + \frac{1}{2}\left(me_2^{(\alpha)}\left(\frac{E_J}{m}\right)^{\alpha}-mz_\text{cut} \right)\Bigl(1+\tanh\bigl( r(e_2^{(\alpha)}-t)\bigr)\Bigr)\,.
\eea
We can verify that for low $e_2^{(\alpha)}$, the profile asymptotes to $mz_\text{cut}$ while for large enough $e_2^{(\alpha)}$ it follows $me_2^{(\alpha)}\left(\frac{E_J}{m}\right)^{\alpha}$  as desired. The parameters $r$ and $t$ are the two knobs which decide the rate of transition and the precise value at which transitions happens respectively. Similarly for the Region I -- fixed-order transition we can choose 
\bea
 \mu_0^{B_+}(e_2^{(\alpha)}) =  me_{2,t}^{(\alpha)}\left(\frac{E_J}{m}\right)^{\alpha}+ \left(E_J-me_{2,t}^{(\alpha)}\left(\frac{E_J}{m}\right)^{\alpha}\right) \tanh \left(r(e_2^{(\alpha)}-e_{2,t}^{(\alpha)})\right)
\eea
where $e_{2,t}^{(\alpha)}$ is the value of $e_2^{(\alpha)}$ at which the profile deviates from its central value in Region I ($me_2^{(\alpha)}\left(\frac{E_J}{m}\right)^{\alpha}$). Once again $r$ and $t$ choose the transition value and the rate of transition between regions. The profile for the collinear-soft function can be tailored in exactly the same way.

\subsection{Comparison with Partonic \textsc{Pythia}}

We present the comparison of our calculations, resummed to NLL accuracy and matched to the one-loop ($\cO(\alpha_s)$) full theory fixed-order cross section, with  \textsc{Pythia} v.~8.219. To make the comparison, we consider parton-level  \textsc{Pythia} results with hadronization and decays turned off.  \fig{bplots} shows the comparison for $b$-quark jets in 2-jet events in $e^+ e^- $ collisions at a center-of-mass energy of $\sqrt{s} = 300$ GeV (left) and 700 GeV (right), with $\alpha = 0.5$ and a $z_\text{cut}=0.1$, using the anti-$k_t$ algorithm with $R=0.6$. \fig{bplots75} shows the same comparison for a CM energy of $\sqrt{s}=150$ GeV. For these jets, $E_J\approx \sqrt{s}/2$, which is the value we also used in our theory predictions. Although these are not typical $\sqrt{s}$ values at any past or existing $e^+e^-$ collider, these choices in \textsc{Pythia} provide us with enough statistics for jets with energies of 50--350 GeV for which we will present results, and which are typical energies of jets at current colliders like LHC or RHIC.
\begin{figure}[b]
\centerline{\scalebox{.45}{\includegraphics{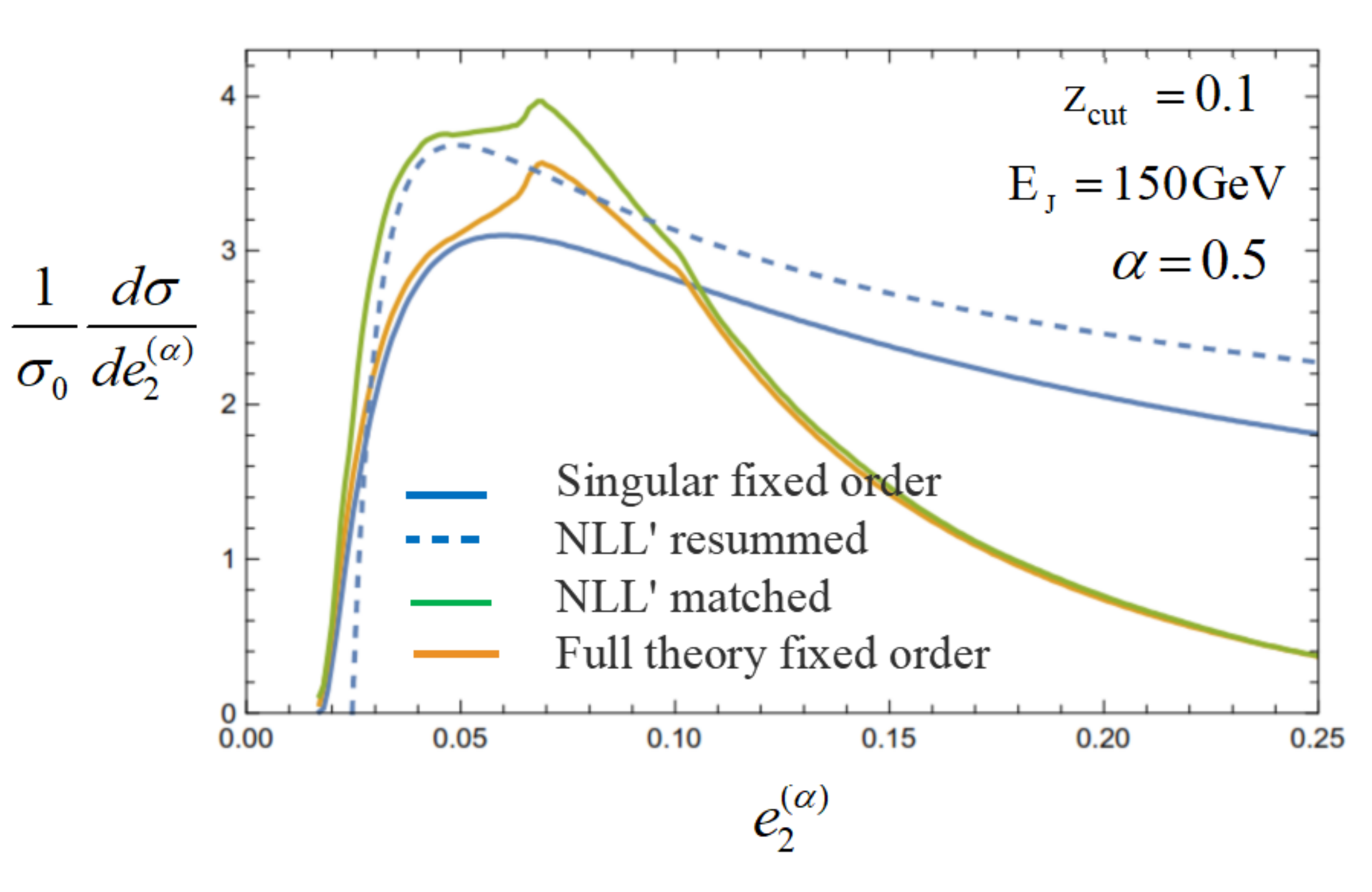}}}
\vskip-0.5cm
\caption[1]{Comparison of resummed and fixed-order cross sections for massive quark jets.}
\label{cmp} 
\end{figure}
\begin{figure}
\centerline{\scalebox{.4}{\includegraphics{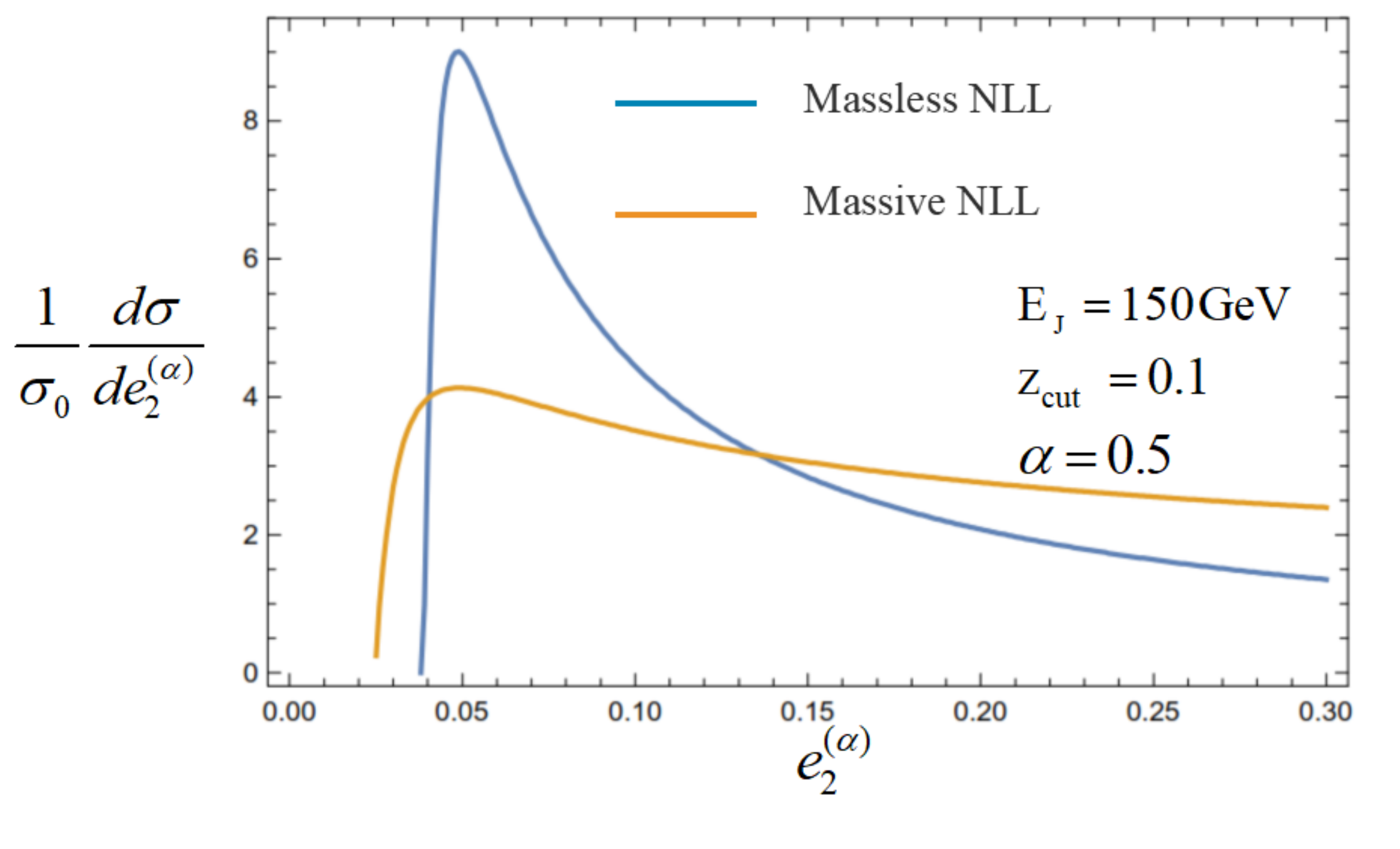}}}
\vskip-0.5cm
\caption[1]{Comparison of resummed perturbative cross section at NLL for the massless and massive quark initiated jets. Note that these curves include no nonperturbative corrections or fixed-order matching.}
\label{fig:masslessComp} 
\end{figure}

To do the matching with the full theory fixed-order result, we use profile functions for the collinear-soft and ultracollinear scales (\fig{profile}) to match the regions I and 0 and also turn off the resummation to smoothly transition to the fixed-order cross section. At the same time, we make an estimate of the errors due to a truncation of the perturbative series by varying these scales by a factor of 2 above and below the central value as shown in \fig{profile}. Each scale is varied independently and the error band generated is added in quadrature to give the final uncertainty estimate. As we see, the error band is of the order of $\pm 10 \%$. To obtain the result at NNLL accuracy, we would need to compute the collinear-soft function at two loops.

We get a good agreement over the whole range of $e_2^{(\alpha)}$. As expected, the cross section goes to zero at a finite non-zero value of $e_2^{(\alpha)}$. We also show the comparison of the fixed-order and resummed cross sections in Fig. \ref{cmp}. As expected, the impact of resummation is most significant in the low $e_2^{(\alpha)}$ region.

We also show a comparison with the resummed, purely perturbative result of the massless quark case at NLL accuracy in \fig{masslessComp}, as a rough estimate of the effect of the quark mass $m_b$ in the resummation region. Since the value of $\alpha$ is small, the non-perturbative corrections set in at a value $e_2^{(\alpha)} \sim 0.05$ for the case of the massless groomed jet. To get a correct picture of the massless spectrum below this value, it is necessary to include non-perturbative corrections. This plot is only meant as a rough illustration of the size of the impact of nonzero $m_b$ in the peak region.


\section{Hadronization and decay effects and comparison with \textsc{Pythia}}
\label{sec:decayq}

\begin{figure}
\centerline{\scalebox{.35}{\includegraphics{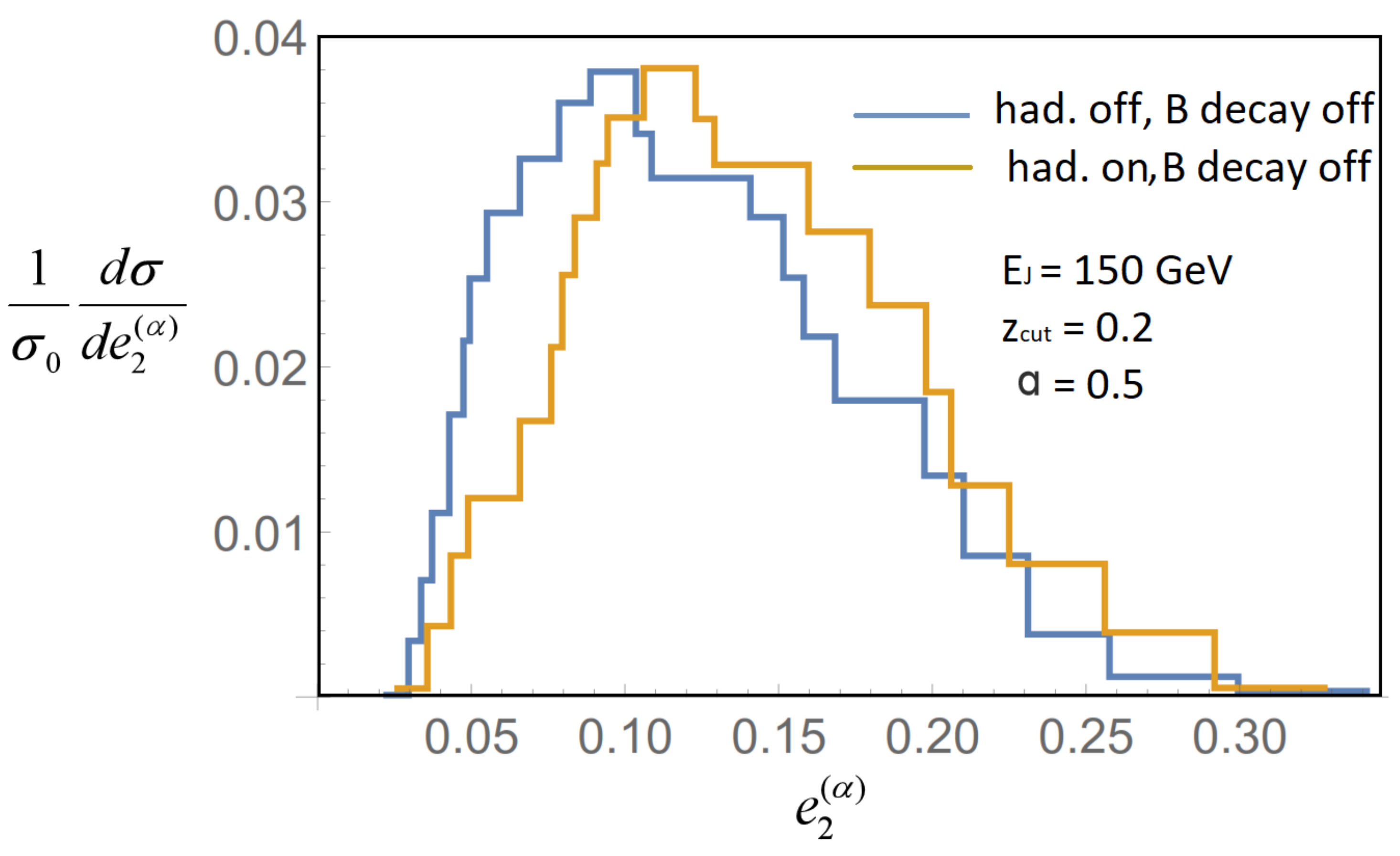}}}
\vskip-0.4cm
\caption[1]{Effect of hadronization (while keeping the $B$ decay turned off)  is to produce a shift in the $e_2^{(\alpha)}$ distribution.}
\label{hadron} 
\end{figure}

We now turn on hadronization in  \textsc{Pythia} while still keeping the $B$ hadron decays off. This enables us to gauge the impact of hadronization corrections. Fig.~\ref{hadron} reveals that the dominant effect of hadronization for large enough $e_2^{(\alpha)}$ is to produce a simple shift in the $e_2^{(\alpha)}$ distribution, much like event shapes and other jet shapes (see, e.g., \cite{Dokshitzer:1995zt,Dokshitzer:1997ew,Dasgupta:2003iq,Lee:2006nr}). This can be understood as the impact of soft radiation ($E \sim \Lqcd$) from outside the jet, hadronizing the colored partons inside the jet to color-singlet bound states. Hence, we expect the energy of the partons to increase by an amount of order $\Lqcd$, thus shifting the distribution towards higher values of $e_2^{(\alpha)}$. The way to incorporate this shift and extract out the nonperturbative physics in terms of measurable parameters is explored in \ssec{hadron}.

\begin{figure}
\centerline{\scalebox{.55}{\includegraphics{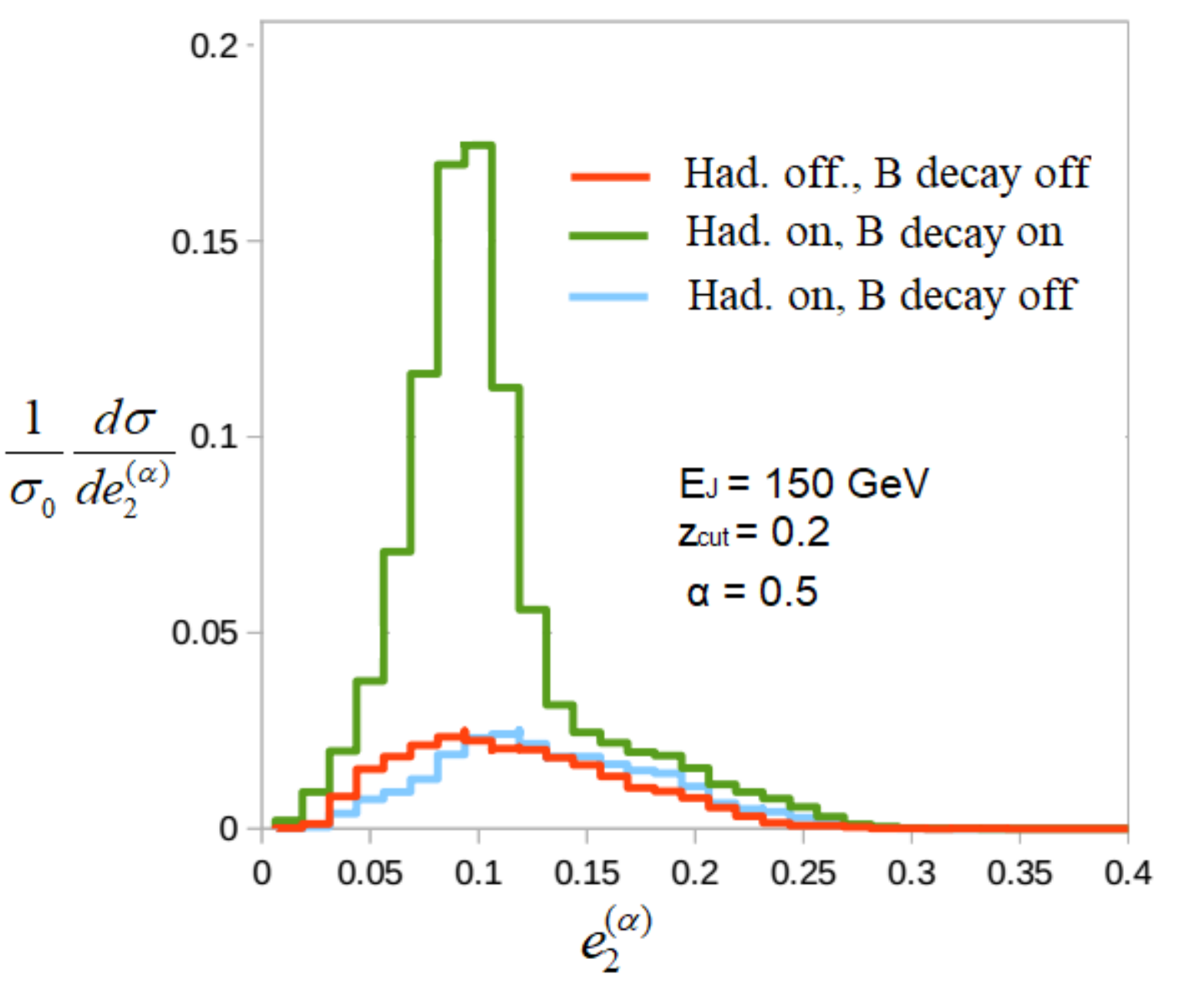}}}
\vskip-0.4cm
\caption[1]{Effect of $B$ hadron decay is to produce a dramatic increase in the number of events contributing to the $e_2^{(\alpha)}$ distribution.}
\label{decay} 
\end{figure}

Next we also turn on the $B$ hadron decay. Comparing it with the purely partonic cross section in Fig.~\ref{decay} shows that the impact of $B$ decay is quite significant. While the peak of the distribution appears to be at a similar value of $e_2^{(\alpha)}$, there is a five-fold increase in the cross section. From Fig.~\ref{decay}, we can see that there are two effects of decay. One simply changes the shape of the original distribution and the other is an addition of a large number of extra events in each $e_2^{(\alpha)}$ bin. 

 \subsection{$B$ decay effects}

The change in the shape is due to change in the radiation pattern for a given event due to the decay of the $B$ hadron. This will change the value of $e_2^{(\alpha)}$ that is being contributed by the partonic radiation configuration. On the other hand, the large number of extra events can be understood as coming from the zero bin of the partonic or the undecayed hadronized distribution. In this bin, at the partonic or undecayed level, we have a single $b$ quark or $B$ hadron in the jet, with all the accompanying radiation groomed away since it fails soft drop. Since this event has only one particle, it will not contribute to a non-zero value of $e_2^{(\alpha)}$. As the $b$ quark hadronizes into a $B$ and subsequently decays (into 2 or more particles), the consequent radiation pattern now has the possibility of contributing to a non-trivial $e_2^{(\alpha)}$. Thus if we have complete information about the branching fraction and the kinematics of the $B$ hadron decay, it should be straightforward to compute how these  zero-bin events change the $e_2^{(\alpha)}$ distribution after decay, even though the effect is large. We leave this computation for future work. This would allow for using $e_2^{(\alpha)}$ itself as a $b$-jet \emph{vs.} light-jet discriminant. (See \cite{Hoang:2017kmk} for inclusion of effects of decay products in top quark jet mass distributions.)

Even though at this stage we only include effects of hadronization but not the decay of the $B$ hadron, we can still make meaningful comparisons with the experiments where the $B$ hadron four-momentum can be reconstructed from its decay products. This can be done using standard techniques for $b$-tagging such as vertex displacement. Then at the level of the reconstructed $B$ hadron, we can measure $e_2^{(\alpha)}$ on the jet, which gives a valuable tool for analyzing the radiation pattern in a jet initiated by a massive quark.


\subsection{Hadronization corrections}
\label{ssec:hadron}

To determine the sector which will give the dominant nonperturbative corrections, we look at the scaling of these corrections for each of our factorized matrix elements, following similar reasoning in \cite{Lee:2006nr}. We first look at the Region I which has both collinear-soft and bHQET functions. 
The measurement function for the collinear-soft function can be written as (Eq.~(\ref{Cmeasure})),
\bea
\label{eq:csmeasurement}
\delta\left( e_2^{(\alpha)}- \frac{1}{2 E_J}  (\bar n \cdot p_i )\left(\frac{n \cdot p_i}{\bar n \cdot p_i } \right)^{\alpha/2}\right) \,.
\eea
To probe for nonperturbative corrections, we reduce the virtuality of this mode to $\Lqcd$ while maintaining the angular scaling $ \theta_{cs} \sim (e_2^{(\alpha)}/z_\text{cut})^{1/\alpha}$.
The non-perturbative collinear-soft mode would then scale as 
\bea
p_{cs} \sim \frac{\Lqcd}{\theta_{cs}}\left(1, \theta^2_{cs}, \theta_{cs}\right) \,.
\eea
We can then deduce the scaling of the leading nonperturbative power correction from the collinear-soft function simply by substituting this scaling for the non-perturbative mode in the expression for the measurement in \eq{csmeasurement}. This gives us for the induced shift in the first moment of the $e_2^{(\alpha)}$ distribution,
\bea
\label{eq:csshift}
\frac{\Delta e_2^{(\alpha)}}{e_2^{(\alpha)}} \sim\frac{\Lqcd}{e_2^{(\alpha)} 2 E_J} \left(\frac{e_2^{(\alpha)}}{z_\text{cut}}\right)^{(\alpha-1)/\alpha} \,.
\eea

For the boosted HQET jet function ($B_+$), the measurement gives (Eq.~(\ref{Hmeasure}))
 \bea
\delta\left( e_2^{(\alpha)} - \frac{\bar n \cdot k}{\omega} \left( 4\frac{ m v_+ \cdot k}{\omega \bar n \cdot k} \right)^{\alpha/2}\right) \,.
\eea
As before, we reduce the virtuality of this mode to $\Lqcd$ while maintaining the angular scaling.
Repeating the logic leading to \eq{csshift}, this gives a leading nonperturbative shift in the $e_2^{(\alpha)}$ distribution that scales as
\bea
\label{eq:bHQETshift}
\frac{\Delta e_2^{(\alpha)}}{e_2^{(\alpha)}}\sim \frac{\Lqcd}{\omega e_2^{(\alpha)}} \left(\frac{\omega}{m_b}\right)^{1-\alpha} \,.
\eea
Recall that $\omega = 2E_J$.

Given the constraint ($e_2^{(\alpha)}>e_{2,\text{min}}^{(\alpha)}$), it is clear that the nonperturbative corrections from the boosted HQET jet function \eq{bHQETshift} will dominate over those from the csoft function \eq{csshift} in the region of measurement, so for the remainder of this analysis we focus only on the bHQET corrections.
Note this also tells us that for the case $\alpha<1$, the leading nonperturbative corrections are independent of grooming. This reasoning also persists in the case of massless jets, where the nonperturbative corrections would come from the SCET jet function in \eq{factorizationmassless}, which is insensitive to grooming. This agrees qualitatively with the picture in \fig{scales}, where we expect the function with the lowest virtuality to contribute to the leading nonperturbative correction in any region. This is also consistent with the implications of the analysis of nonperturbative corrections in \cite{Larkoski:2014wba}.

Following the discussion in \cite{Lee:2006nr,Lee:2006fn}, using the definition of rapidity as $ \eta = \frac{1}{2} \ln (\bar n \cdot k/n \cdot k)$, we can rewrite the  measurement delta function for the bHQET jet function as
\bea
 \delta\left( e_2^{(\alpha)} -\frac{\bar n \cdot k}{\omega} \left( 4\frac{ m v_+ \cdot k}{\omega \bar n \cdot k} \right)^{\alpha/2}\right) =\delta\left( e_2^{(\alpha)} - \frac{k_{\perp}}{\omega} e^{\eta}\left( 4(\Delta+ e^{-2\eta}) \right)^{\alpha/2}\right) \,.
\eea 
Let us come back to the definition of the bHQET jet function, 
\bea
B_+(e_2^{(\alpha)},m) = \langle 0 | \bar h_{v_+} W_n \delta\left(e_2^{(\alpha)} -  \frac{k_{\perp}}{\omega} e^{\eta}\left( 4(\Delta+ e^{-2\eta}) \right)^{\alpha/2}\right) W_n^{\dagger} h_{v_+} |0 \rangle,
\eea
and transform it to Laplace space $(s)$, $\widetilde B_+(s, m) = \int d e_2^{(\alpha)} e^{-s e_2^{(\alpha)}} B_+(e_2^{(\alpha)},m)$, which gives
\bea
 \widetilde B_+(s, m) =  \langle 0 | \bar h_{v_+} W_n e^{-  \frac{s k_{\perp}}{\omega} e^{\eta}\left( 4(\Delta+ e^{-2\eta}) \right)^{\alpha/2}} W_n^{\dagger} h_{v_+} |0 \rangle.
\eea
If we consider a regime where $ k_{\perp} \sim \Lqcd$, then we can consider the exponent as a power series in $ \Lqcd/\Gamma$ where $\Gamma$ is defined in Eq.~(\ref{gamma}). The dominant power correction will then be captured by the leading non-trivial term of this series,  
\bea
\label{eq:P1}
 P_1= - \sum_X \langle 0 | \bar h_{v_+} W_n \left( \frac{s k_{\perp}}{\omega} e^{\eta}\left( 4(\Delta+ e^{-2\eta}) \right)^{\alpha/2}\right)|B+X\rangle \langle B+X| W_n^{\dagger} h_{v_+} |0 \rangle.
\eea
Following \cite{Lee:2006nr,Lee:2006fn}, we can write this in terms of the energy flow operator \cite{Ore:1979ry,Sveshnikov:1995vi,Korchemsky:1997sy,Korchemsky:1999kt,Belitsky:2001ij,Bauer:2008dt} as
\bea
 P_1=  \sum_X \int d \eta f(\eta)  \langle 0 | \bar h_{v_+} W_n \mathcal{E}_T(\eta)|B+X\rangle \langle B+X| W_n^{\dagger} h_{v_+} |0 \rangle,
\eea
where the operator $\mathcal{E}_T(\eta)$ gives us 
\bea
 \mathcal{E}_T(\eta)|X \rangle = \sum_{i\in X} \bigl\lvert\vect{k}_i^\perp\bigr\rvert \delta(\eta- \eta_i)|X\rangle, 
\eea
which then defines $f(\eta)$ as 
\bea
f(\eta) = -s e^{\eta}\left( 4(\Delta+ e^{-2\eta}) \right)^{\alpha/2}.
\eea
Unlike for soft power corrections in \cite{Lee:2006nr,Lee:2006fn}, we cannot completely disentangle the $\eta$ dependence of $f$ and the matrix element in \eq{P1}, which otherwise would determine $P_1$ in terms of a single universal nonperturbative matrix element. However we can still extract the scaling of $P_1$ in $m$ and $\Delta$ using similar manipulations as in \cite{Lee:2006nr,Lee:2006fn}. 

In order to do this, we can boost all the operators to the rest frame of the heavy quark. This is basically a boost in the $-z$ direction. We know that the collinear Wilson line $W_n$ remains invariant under this boost as a result of the RPI III symmetry of SCET \cite{Manohar:2002fd}. The vacuum state is invariant under Lorentz transformations. Under our boost, the field $h_{v_+}$ transforms to $h_{\tilde v}$ where $\tilde v =(1,1,0)$, i.e., the rest frame of the heavy quark. Finally, the energy flow operator transforms as 
\bea
\mathcal{E}_T(\eta) \rightarrow  \mathcal{E}_T(\eta+\eta')
\eea
where $ e^{\eta'} = (m/E_J)$. We then have 
\bea
P_1(s) = \sum_X\int d \eta f(\eta)  \langle 0 | \bar h_{\tilde v} W_n \mathcal{E}_T(\eta+\eta')|B+X\rangle \langle B+X| W_n^{\dagger} h_{\tilde v} |0 \rangle.
\eea
Redefining $\eta$ as $\eta+\eta'$, we are now left with 
\bea 
P_1(s) =\sum_X \int d \eta f(\eta-\eta')  \langle 0 | \bar h_{\tilde v} W_n \mathcal{E}_T(\eta)|B+X\rangle \langle B+X| W_n^{\dagger} h_{\tilde v} |0 \rangle,
\eea
which simplifies to
\bea
P_1 (s)= -\frac{s (4\Delta)^{\alpha/2}}{m} \sum_X\int d \eta F(\eta, \alpha )  \langle 0 | \bar h_{\tilde v} W_n \mathcal{E}_T(\eta)|B+X\rangle \langle B+X| W_n^{\dagger} h_{\tilde v} |0 \rangle,
\eea
where 
\bea
F(\eta,\alpha) = e^{\eta} \left(1+ e^{-2\eta}\right)^{\alpha/2},
\eea
which is now independent of both the mass and energy, though it still depends on $\alpha$ and $z_{\text{cut}}$.
Note that, plugging in $\Delta = m^2/\omega^2$, counting the matrix element of $\mathcal{E}_T$ in the nonperturbative regime as $\Lqcd$, and counting the rapidity integral of $F$ as $\cO(1)$ for soft radiation in the $B$ rest frame in which this expression is evaluated, the power counting for this leading nonperturbative correction is consistent with our initial estimate in \eq{bHQETshift}.

\begin{figure}
\centerline{\scalebox{.35}{\includegraphics{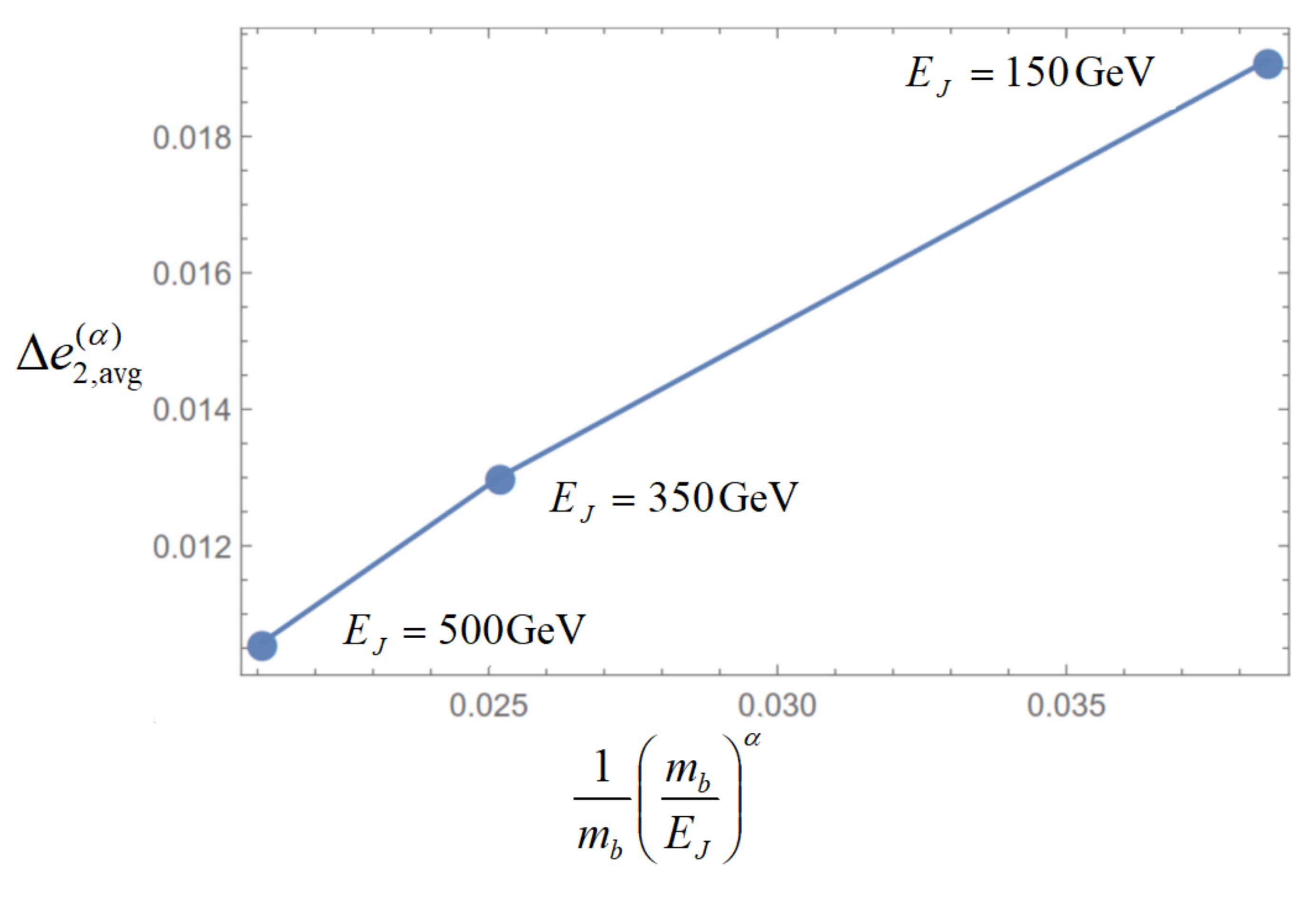}}}
\vskip-0.5cm
\caption[1]{The slope of the curve gives us a value of $A(\alpha)=0.5$ GeV for $\alpha=0.5$}.
\label{fig:nonpert} 
\end{figure}
In Laplace space we can write the leading power correction as 
\bea
\frac{1}{\sigma_0} \frac{d\sigma}{d e_2^{(\alpha)}} =   \sigma(e_2^{(\alpha)} )_{\text{perturbative}} - \frac{ (4\Delta)^{\alpha/2}}{m}A(\alpha)\frac{d}{de_2^{(\alpha)}} \sigma(e_2^{(\alpha)} )_{\text{perturbative}}
\eea
where 
\bea
A(\alpha) = \sum_X \int d \eta F(\eta, \alpha )  \langle 0 | \bar h_{\tilde v} W_n \mathcal{E}_T(\eta)|B+X\rangle \langle B+X| W_n^{\dagger} h_{\tilde v} |0 \rangle, 
\eea
which is an $\alpha$ dependent (but independent of the mass and energy) nonperturbative parameter which we need to extract from simulation/experiment. The extraction can be done at a single  arbitrary energy, mass and then can be used for any other energy and mass. In order to extract this parameter, we consider the first moment of the cross section.  Assuming the perturbative distribution is normalized so that the area under the perturbative differential distribution is $1$, we can then immediately write
\bea
(e_2^{(\alpha)})^{\text{avg}} = (e_2^{(\alpha)})^{\text{avg}}_{\text{perturbative}}+   \frac{ (4\Delta)^{\alpha/2}}{m}A(\alpha).
\eea
Given a specific value of $\alpha$, we can predict the scaling of the shift in the average value of $e_2^{(\alpha)}$ with the mass of the heavy quark and the jet energy.
For the case of the $b$ quark at $\alpha=0.5$, from a rough estimate from \fig{nonpert}, we get a value of $A(\alpha) =0.5$ GeV. 

In the Region 0 of factorization, there is a single function which is the bHQET jet function with a soft drop constraint ($B_+^{SD}$). The analysis for the hadronization correction in this region follows in the same manner as the before. However, since this region covers only as small range of $e_2^{(\alpha)}$ and exists near $e^{(\alpha)}_{2,\text{min}}$, the contribution to the average value of $e_2^{(\alpha)}$ and hence $A(\alpha)$ is very small. Hence, to a very good approximation, the value of $A(\alpha)$ is entirely determined by from the region I.


\section{Conclusion}
\label{sec:conclusion}

In this paper, we have presented the first analytic EFT resummation for energy correlator distrbutions computed on groomed massive jets, subsequently matched to numerical fixed-order $\cO(\as)$ results. 
We begin with a calculation of the full theory fixed-order cross section for the correlator $e_2^{(\alpha)} $ for a jet initiated by a massive quark.
The fixed-order cross section has several interesting features such as a minimum value for the observable, along with kinks at values of $e_2^{(\alpha)}$ where grooming and quark mass effects kick in.  We provide an intuitive understanding of these features in terms of specific physical configuration of the radiation in the jet. This analysis helps us understand the relevant physical scales in this process and hence the scaling of the radiation in a given range of $e_2^{(\alpha)}$. The minimum value of $e_2^{(\alpha)}$ is enforced by the mass of the quark and the soft-drop energy cut-off. We conclude that the mass of the quark is relevant in the regime $e^{(\alpha)}_{2,\text{min}} < e_2^{(\alpha)} < (m/E_J)^{\alpha}$ (``Region I''), while the system behaves as a massless quark groomed jet in the region $(m/E_J)^{\alpha} < e_2^{(\alpha)} < z_\text{cut}$ (``Region II'') up to power corrections in $m^2/E^2_J$. In the tail region, $e_2^{(\alpha)} > z_\text{cut}$, the  distribution is essentially that of a massless quark ungroomed jet. Depending on the size of $\alpha$, we may or may not need a separate Region II. In the present work we primarily explored the phenomenology of setup ($\alpha<1$) needing only Region I.

To that end we utilize an EFT framework combining the SCET$_+$ and bHQET formalisms for the heavy quark and the radiation it emits, which is valid for $e^{(\alpha)}_{2,\text{min}} < e_2^{(\alpha)} < (m/E_J)^{\alpha}$, i.e. Region I. In cases where a Region II EFT is needed ($\alpha>1$), we discussed the need to make a transition to the massless quark EFT regime which has been developed in \cite{Frye:2016aiz}, again within the SCET$_+$ formalism. We observe that the two EFT descriptions are identical at the boundary value $ e_2^{(\alpha)} = (m/E_J)^{\alpha}$, so that one EFT naturally transitions into the other at this value For our case of interest, $\alpha<1$, we transition directly from Region I to the fixed-order regime. At the same time, we observe that the EFT for the region $e^{(\alpha)}_{2,\text{min}} < e_2^{(\alpha)} < (m/E_J)^{\alpha}$ can  be further divided up into two regions, one very near the endpoint $e_{2,\text{min}}^{(\alpha)}$ (``Region 0''), and then the normal Region I once we move sufficiently above $e_{2,\text{min}}^{(\alpha)}$, where corrections of order $e^{(\alpha)}_{2,\text{min}} / e_2^{(\alpha)}$ are power suppressed.

We compute the anomalous dimension for these EFTs and resum the large logarithms in $e_2^{(\alpha)}$ to NLL$'$ accuracy.  The resummed cross section is matched onto the one loop $\mathcal{O}(\alpha_s)$ full theory fixed-order cross section. The result shows good agreement with partonic  \textsc{Pythia} simulation results. Comparing with hadronic  \textsc{Pythia} (with $B$ hadron decays turned off) shows a simple shift in the distribution. We then use the energy flow operator formalism to calculate the scaling of the leading nonperturbative correction to the cross section, which can describe this shift in the distribution. An unusual feature of this distribution is that the dominant nonperturbative corrections appear in the bHQET jet function (as opposed to the soft function for a massless jet), for the values of $z_\text{cut}$ and $\alpha$ that we considered. Using the Lorentz transformation properties of the bHQET jet function, we can extract out a universal $\alpha$ dependent nonperturbative parameter (independent of mass and energy), which then can be extracted from simulations/experiment.
 
The distribution we have computed gives an accurate prediction at the level of $B$ hadron production, before its decay. In other words, to compare with experiment, it is necessary to reconstruct the momentum of the $B$ hadron from its decay products and then compute the observable using the $B$ hadron four-momentum. This can, in principle be achieved using $b$ tagging techniques such as vertex displacement. 
Turning on $B$ decay produces a significant deviation from the distribution that we computed in this way, primarily due to a large number of extra events which populate nonzero $e_2^{(\alpha)}$ bins. 
Including such effects in our calculation will greatly expand the applicability of this observable as a probe of heavy quark jet substructure. However, this is something we leave for the future.

In order to extend the accuracy of our results to NNLL accuracy, the only additional ingredient needed is the two loop collinear-soft function. At present, this function is only known for the special case of $\alpha =2$, which corresponds to jet mass. Even though we have considered only the two particle correlator, it is clear that the general form of the EFT developed  in this paper and the observations made about the relevance of the quark mass will be applicable for other jet shape observables computed on heavy quark initiated jets.

We believe that this is a significant step into a much more detailed study of jet substructure for heavy quark jets. This analysis will serve as a template for comparison with the corresponding cross section for heavy quark jets in a medium (QGP), which should be able to reveal the effects of medium on jet substructure (see e.g. \cite{Li:2017wwc}).

\begin{acknowledgments}
 This work was supported by the U.S. Department of Energy through the Office of Science, Office of Nuclear Physics under Contract DE-AC52-06NA25396 and by an Early Career Research Award, through the LANL/LDRD Program, and within the framework of the TMD Topical Collaboration. P.S. was supported by the DOE contracts DE-FG02-04ER41338 and DE-FG02-06ER41449. 
\end{acknowledgments}
\appendix

\bibliography{bjet}{}

\end{document}